\documentclass{article}

\usepackage{algorithm}
\usepackage{algorithmic}

\usepackage{tabularx}

\usepackage{adjustbox}
\usepackage{arxiv}
\usepackage[utf8]{inputenc} 
\usepackage[T1]{fontenc}    
\usepackage[colorlinks=true]{hyperref}  
\usepackage[utf8]{inputenc} 
\usepackage[T1]{fontenc}    
\usepackage{hyperref}       
\usepackage{url}            
\usepackage{booktabs}       
\usepackage{amsfonts}       
\usepackage{nicefrac}       
\usepackage{microtype}      
\usepackage{fancyhdr}       
\usepackage{lipsum}
\usepackage{natbib}
\usepackage{fancyhdr}       
\usepackage{graphicx}       
\graphicspath{{media/}}     
\usepackage{url}            
\usepackage{booktabs}       
\usepackage{amsfonts}       
\usepackage{nicefrac}       
\usepackage{microtype}      
\usepackage{fancyhdr}       
\usepackage{graphicx}       
\graphicspath{{media/}}     
\usepackage{amssymb}
\usepackage{amsmath}
\usepackage{mathtools}
\usepackage{placeins}
\usepackage{natbib}
\usepackage{caption}
\usepackage{verbatim}
\usepackage{rotating}
\usepackage[utf8]{inputenc}
\usepackage{pgfplots}
\usepackage{appendix}
\usepgfplotslibrary{groupplots}
\newcommand{\vx}{\boldsymbol{x}}

\newcommand{\vu}{\boldsymbol{u}}

\newcommand{\vy}{\boldsymbol{y}}

\newcommand{\cm}{\mathcal{M}}

\newcommand{\ve}[1]{\boldsymbol{#1}}

\renewcommand{\shorttitle}{Bayesian full waveform inversion with Surrogate Modeling }

\title{Bayesian full waveform inversion with sequential surrogate model refinement}

\author{
  Giovanni Angelo Meles\\
    Nederlandse Organisatie voor  \\
      Toegepast Natuurwetenschappelijk Onderzoek \\
      TNO \\
      The Netherlands\\
    \texttt{Giovanni.Meles@tno.nl} \\
  \And
  Stefano Marelli \\
  Institute of Structural Engineering \\
  ETH Zurich \\
  Switzerland\\
  \texttt{marelli@ibk.baug.ethz.ch} \\
       \And
     Niklas Linde \\
    Institute of Earth Sciences \\
      University of Lausanne \\
      Switzerland\\
    \texttt{Niklas.Linde@unil.ch} \\
}

\begin{document}
\maketitle
\begin{abstract}

Bayesian formulations of inverse problems are attractive due to their ability to incorporate prior knowledge, account for various sources of uncertainties, and update probabilistic models as new information becomes available.
Markov chain Monte Carlo (MCMC) methods sample posterior probability density functions (pdfs) provided accurate representations of prior information and many evaluations of likelihood functions. Dimensionality-reduction techniques such as principal component analysis (PCA) can assist in defining the prior pdf and the input bases can be used to train surrogate models. Surrogate models offer efficient approximations of likelihood functions that can replace traditional and costly forward solvers in MCMC inversions. Many problem classes in geophysics involve intricate input/output relationships that conventional surrogate models, constructed using samples drawn from the prior pdf fail to capture, leading to biased inversion results and poor uncertainty quantification. Incorporating samples from regions of high posterior probability in the training may increase accuracy, but identifying these regions is challenging.
In the context of full waveform inversion, we identify and explore high-probability posterior regions using a series of successively-trained surrogate models covering progressively expanding wave bandwidths. 
The initial surrogate model is used to invert low-frequency data only as the input/output relationship of high-frequency data are too complex to be described across the full prior pdf with a single surrogate model. 
After a first MCMC inversion, we retrain the surrogate model on samples from the resulting posterior pdf and repeat the process. By focusing on progressively narrower input domain regions, it is possible to progressively increase the frequency bandwidth of the data to be modeled while also decreasing model errors. Through this iterative scheme, we eventually obtain a surrogate model that is of high accuracy for model realizations exhibiting significant posterior probabilities across the full bandwidth of interest. This surrogate model is then used to perform an MCMC inversion yielding the final estimation of the posterior pdf. Numerical results from 2D synthetic crosshole Ground Penetrating Radar (GPR) examples demonstrate that our method outperforms ray-based approaches, as well as results obtained when only training the surrogate model using samples from the prior pdf. Our methodology reduces the overall computational cost by approximately two orders of magnitude compared to using a classical finite-difference time-domain forward scheme.
\end{abstract}
\keywords{Inverse theory \and full waveform inversion \and modeling refinement \and surrogate modeling  \and Bayesian statistics}

\section{Introduction}

Tomography encompasses a diverse array of non-invasive imaging methods that are widely used across numerous application fields, such as exploration for natural resources \citep{taillandier2009first}, investigating moons and other planets \citep{zhao2008seismic,khan2016single}, diagnosing medical conditions \citep{stotzka2002medical}, and conducting non-destructive testing \citep{tant2018transdimensional}. Full Waveform Inversion (FWI) stands out given its potential to maximize information retrieval across multiple scales. 
Advancements in computational capabilities have facilitated FWI developments \citep{dhabaria2024hamiltonian}, leading to its growing use for characterizing geological structures, assessing natural resources, and mitigating environmental risks \citep{ernst2007application,mulder2008exploring,Virieux2009,fichtner2013multiscale}.
Deterministic FWI can generate images that reveal intricate geological features and resolve material properties at sub-wavelength scale. However, the associated algorithms are prone to a number of limitations, particularly with regard to their limited capabilities for uncertainty quantification and the high risk of converging to a local minimum \citep{virieux2017introduction,klotzsche2019review}.
In contrast, Bayesian inversion methods treat model parameters as random variables can avoid such issues \citep{liu2001monte,ulrych2001bayes}. Nevertheless, the substantial computational demand stemming from the need to perform many forward waveform simulations can be exceedingly demanding. Various strategies have been employed to enhance convergence and mitigate computational expenses of Bayesian FWI. These strategies includeadvanced MCMC schemes inspired by evolutionary concepts, Hamiltonian Monte Carlo (HMC) and Variational Bayesian inference (VBI) \citep{hunziker2019bayesian,gebraad2020bayesian,aleardi2020combining,zhang20233}. 

An increasingly common set of tools to reduce computational costs in uncertainty quantification is surrogate modelling, such as Kriging \citep{sacks1989designs}, polynomial chaos expansions (PCEs) \citep{xiu2002wiener,luthen2022automatic, meles2022bayesian} or deep neural networks (DNN) methods \citep{adler2018deep, wang2018velocity,wu2018inversionnet,yang2019deep,sun2023implicit}.
Developing accurate surrogate models for physical systems often requires reducing the dimensionality or simplifying the representation of input variables to effectively approximate complex system behaviors. Successful surrogate modeling will enhance computational feasibility while preserving a high level of accuracy in input representation and output prediction.
Various approaches exists to achieve effective dimensionality reduction, with principal component analysis (PCA) being the most common \citep{boutsidis2008unsupervised,jolliffe2016principal,reynolds1996reparameterization,giannakis2021fractal,meles2022bayesian,thibaut2021new}.

Recently, \citet{meles2022bayesian} performed Bayesian inversion with a PCA-based parameterization of prior information to predict travel times with PCE. Extending this strategy to full-waveform inversion is non-trivial due to more intricate input-output relationships compared to traveltime modeling \citep{meles2024bayesian} and the limited expressiveness provided by PCE. This limitation arises not from theoretical arguments but from the truncation schemes that are needed to limit the number of polynomial coefficients to be evaluated.

To address the Bayesian FWI problem, we introduce a sequential formulation \citep{peherstorfer2018survey} involving surrogate models that progressively cover expanding data bandwidths, trained on progressively smaller manifolds. 
This approach leverages heuristic arguments about input-output relationships in wave phenomena, focusing initially on constructing effective surrogate models for low-frequency data components over the whole prior probability density function (pdf). By iteratively refining model training within narrower input regions (i.e., drawn from intermediate posteriors obtained by inverting lower-frequency data), our method aims to achieve accurate modeling across the full frequency bandwidth of interest over the support of the posterior pdf to enable accurate and efficient inversion of all data. 

\section{Methodology}

\subsection{Bayesian inversion via surrogate modeling}
\label{sBayes}
Numerical forward models are used to predict the outcomes of physical experiments for given values of input parameters. They describe the relationship between input parameters and the output as:
\begin{equation}
 \mathcal{F}(\vu) = \vy + \epsilon.
 \label{forward}
\end{equation}
Here, $\mathcal{F}$ represents the physical law or forward operator, typically involving partial differential equations (PDEs) acting on locally defined parameters. The variable $\vu$ denotes the input parameters, $\vy$ the data, and $\epsilon$ is a noise term. The inverse problem seeks to infer properties of $\vu$ given the data $\vy$ with a prescribed noise model of the data acquisition process while considering any available prior information about $\vu$.

For a constant model input dimension, a general formulation of this problem can be expressed through the unnormalized posterior pdf:
\begin{equation}
 P(\vu|\vy)\propto P(\vy|\vu)P(\vu)
 \label{Bayes}=L(\vu)P(\vu).
\end{equation}
In this context, $P(\vu|\vy)$ represents the posterior pdf of the input $\vu$ given the \textit{n}-dimensional data $\vy$. The probability $P(\vy|\vu)$, also denoted as $L(\vu)$ and referred to as 'the likelihood,' is the probability of observing the data $\vy$ given the input parameter values $\vu$. Finally, $P(\vu)$ is the prior pdf of the input parameters. In this paper, we follow standard formalism in geophysics and use the same symbol (e.g., $\vu$) to denote both individual instances of a random variable and the random variable itself \citep{tarantola2005inverse, aster2018parameter}. 

\textcolor{black}{For zero-mean Gaussian observational noise and a perfect forward operator, the likelihood takes the following form:
      \begin{equation} 
            \label{ExactLikelihood}
   L(\vu)=   \left( \frac{1}{2 \pi} \right)^{n / 2} |\boldsymbol{C_d}|^{-1/2} \mbox{exp} 
           \left[ -\frac{1}{2} (\mathcal{{F}}{(\vu)}  - {\vy} )^T {\boldsymbol{C_d}}^{-1} (\mathcal{{F}}{(\vu)}  - {\vy}) \right] \,
\end{equation}
where the data covariance matrix $\boldsymbol{C}_d$ quantifies  data uncertainty.
To characterize the unnormalized posterior pdf in Eq. ~\eqref{Bayes}, it is common to use Markov chain Monte Carlo (MCMC) methods to draw samples proportionally from $P(\vu|\vy)$ \citep{gelman1995bayesian}.} 

Computing the likelihood $L(\vu)$ typically involves using physics-based solvers, which can be computationally demanding for Bayesian inversion that rely on very large numbers of likelihood computations. In recent years, surrogate modeling has gained popularity as an alternative to physics-based solvers, even if its use in geophysics has been limited until recently \citep{LINDE2017166}. 
Surrogate modeling offers an approximate and computationally inexpensive representation of the forward model. 
The coordinate system used to express the forward model in Eq.\eqref{forward} may not be suitable for implementing a surrogate model, particularly if the dimensionality of the input $\vu$ is  high. In many cases, the forward model may nonetheless be effectively represented in a lower-dimensional manifold, enabling a proper surrogate formulation. 
\label{effective}

To achieve this, one needs to express the forward problem in terms of a new set of variables. Among various approaches, PCA is particularly effective for this task, as it optimally captures input variability through a linear representation, enabling accurate low-dimensional approximations.
The input $\vu$ in Eq.~\eqref{forward} can be uniquely represented as $\vx_{full}$ by projecting it onto a complete, possibly high-dimensional, set of principal components, which preserves the input-output relationship without introducing any approximation.
\begin{equation}
 \mathcal{F}(\vu({\vx_{full}})) = \mathcal{M}({\vx_{full}}) = {\vy} +\epsilon,
 \label{sforward}
\end{equation}
where $\mathcal{M}=\mathcal{F\circ\vu}$ with "$\circ$" indicating function composition. 
It is then possible to truncate the PCA expansion and retain only the first $M$ principal components. In the following, we denote by $\vx$ the $M$-dimensional approximation of $\vx_{full}$.
The forward model can be accurately approximated using a reduced $M$-dimensional representation $\vx$, whenever the discrepancy between the model output computed on $\vx_{full}$ and on its truncated counterpart $\vx$ is negligible. When $\mathcal{M}(\vx_{full}) \approx \mathcal{M}(\vx)$, we can then write:
\begin{equation}
 \mathcal{M}({\vx}) = {\vy} +\hat{\epsilon},
 \label{forwardreduced}
\end{equation}
where $\hat{\epsilon}$ encompasses both observational noise and modeling errors due to the truncation with the truncation level 
$M$ dictating the accuracy in representing the input domain. The required truncation needed to achieve the desired accuracy in modeling depend on the forward problem. This is because non-linear modeling non-trivially propagates the inaccuracies from the input to the output, making it impossible to determine $M$ a priori. Thus, different forward models may require different levels $M$ of PCA truncation to ensure satisfactory results. 
Operating on an effective set of coordinate ${\vx}$ casts the inverse problem on the $M$-dimensional manifold as:
\begin{equation}
 P(\vx|\vy)\propto P(\vy|\vx)P(\vx)
 \label{BayesReduced}=L(\vx)P(\vx).
\end{equation}
Once the forward problem and Bayes' theorem are expressed in terms of a low-dimensional input, surrogate modeling can be used for MCMC \citep{chib1995understanding} scheme, with the reference solver $\mathcal{M}(\vx_{full})$ being replaced by a surrogate $\tilde{\mathcal{M}}(\vx)$,:
\begin{equation}
\tilde{\mathcal{M}} (\vx ) \approx \mathcal{M} (\vx_{full}).
 \label{surrogate}
\end{equation}
For likelihood evaluation in MCMC algorithms, surrogate models should be employed with a modified  covariance operator $\boldsymbol{C}_D = \boldsymbol{C}_d+\boldsymbol{C}_{Tapp}$ with $\boldsymbol{C}_{Tapp}$ accounting for the impact of truncation and surrogate modeling errors: 
      \begin{equation} 
      \label{likely}
   L(\tilde{M}(\vx ))=   \left( \frac{1}{2 \pi} \right)^{n / 2} |\boldsymbol{C_D}|^{-1/2} \mbox{exp} 
           \left[ -\frac{1}{2} (\mathcal{\tilde{M}} {(\vx )}  - {\vy} )^T {\boldsymbol{C_D}}^{-1} (\mathcal{\tilde{M}} {(\vx )}  - {\vy}) \right] \,
\end{equation}
where $|\boldsymbol{C_D}|$ is the determinant of the covariance matrix $\boldsymbol{C_D}$ \citep{hansen2014accounting}.


Since the computational cost of evaluating a surrogate model typically scales with the output dimension, reducing the number of output variables can significantly accelerate execution. When the output exhibits strong correlations or redundancy, techniques such as PCA can be leveraged to construct a more compact yet informative representation, improving efficiency without compromising predictive capability \citep{meles2022bayesian}. In the context of waveform data, PCA can, in principle, be applied to full-bandwidth time-domain signals. In the following sections, we discuss the challenges of directly applying surrogate modeling to time-domain data and propose a potential workaround that combines monochromatic filtering with PCA decomposition of the output to enable an effective implementation.

\subsection{Surrogate modeling of wave-based phenomena}

Common surrogate modeling approaches such as Kriging \citep{sacks1989designs} or polynomial chaos expansions \citep{xiu2002wiener}, are typically designed to approximate scalar-output models. 
The simplest extension to models with multivariate outputs is based on separately modeling each output variable, sometimes coupled with dimensionality reduction techniques such as PCA \citep{blatman2011principal}). Some specific classes of surrogates, such as Kriging, also admit the explicit inclusion of covariance information, albeit sometimes at the cost of an overall reduced accuracy \citep{KLEIJNEN2014573}.

Surrogate modeling of systems with time dependent outputs of the form $y(\vx,t) = \cm(\vx,t)$, however, is generally a much more complex task. 
This is due to both the output vector dimension, often discretized in the order of $O(10^{2-5})$ timesteps, but mainly because of the increase in prevalence of higher order effects at later times, due to interferences, reflections, etc..
We make an important distinction between two main families of dynamical systems, based entirely on the characteristics of the input excitation. 
We define the first class as \textit{systems with fundamentally simple inputs}, the input of which are either not time-dependent (e.g. Green's functions), or exhibit variability that can be accurately represented with a relatively small number of scalar parameters (e.g. monochromatic waves, fixed wavelets with variable amplitude and dominant frequency, etc.). 
This class of systems can generally be treated with classical surrogate modeling approaches, combined with advanced dimensionality reduction techniques, for example, those presented in \cite{ChuTimeWarping2017,yaghoubi2017sparse,meles2022bayesian}.

The second class is that of \textit{systems with fundamentally complex inputs}, characterized by input excitations that are poorly compressible, that is, their variability requires a large to infinite number of parameters to be properly represented. 
Typical input excitations of this kind are turbulent winds and earthquakes. 
Approximating this class of systems is much more challenging, and it is the territory of state-space models, such as autoregressive models with exogenous inputs (NARX, \cite{billings_2013}), and their more recent incarnations \citep{schar2024emulating}.

In the case of waveform tomography, the system belongs to the first class of models. 
Despite the very high output dimensionality (each realization of the input model $\ve x$ can result in  large set of output traces, typically in the order of $O(10^2)$ traces, each counting $O(10^3)$ timesteps), the set of input wavelets (one per source position) is considered constant.
Therefore, even if the overall problem dimensionality can be large, due to the choice of subsurface parametrization, the time dependence of the input excitations is very low dimensional.
This peculiarity of the tomographic inverse problem allows us to focus our efforts onto devising effective output dimensionality reduction strategies, while still deploying powerful strong learners from the UQ literature, rather than developing our own problem-specific surrogate model to solve both.

Surrogates typically train on samples drawn from the prior pdf. This generally works well if the prior pdf is not too wide, if there are not too many input and output dimensions and if input-output relationships are not highly non-linear \citep{meles2022bayesian}. However, the strategy by \citet{meles2022bayesian} is likely to fall short when applied to non-linear waveform modeling across prior pdfs with wide support. An alternative training strategy is to focus on the much smaller hyper-volume of the parameter space exhibiting significant posterior probability densities \citep{cui2015data}. These regions can be located and sampled through an initial inversion using a reasonably accurate surrogate model trained on samples from the prior pdf that in turn is used for a new MCMC inversion. Such sequential approach involving surrogates of different fidelities can often produce more accurate inference than using one fidelity-level only while still maintaining important speed-ups compared with using the original computationally-costly forward solver only \citep{peherstorfer2018survey, Amaya_etal_2024}.  For non-linear problems, constructing such an initial surrogate model of sufficient accuracy can be challenging if considering all data. 
An alternative is to first restrict the inversion to the regions of the input and output space where the initialy surrogate model is sufficiently accurate. 


For the considered Bayesian FWI problem, we leverage knowledge about wave phenomena to guide the initial selection of data dimensions and input parameters to be used. Our strategy considers the well-established fact that low-frequency data primarily reflects low-wavenumber structures in the velocity distribution while higher-wavenumber features, corresponding to smaller and more abrupt spatial variations, have a significant impact on higher-frequency waves \citep{williamson1990tomographic,bunks1995multiscale,chew1995frequency,pratt1998gauss,zhou2003crosshole,Meles2012a}.
The complexity of wave propagation increases with frequency due to the shorter wavelengths involved, thereby, making them more sensitive to small-scale heterogeneities. This sensitivity leads to intricate interference patterns, multipath effects, and strong attenuation and dispersion, contrasting with the more uniform and predictable propagation of low-frequency waves with longer wavelengths \citep{Aki2002}.
Consequently, we expect that low-frequency data can be modeled far more reliably with surrogate modeling than high-frequency data across the entire prior pdf, while accurate modeling of high-frequency data requires training on a more restricted subset of the input space.

In the following, we present two algorithms for waveform data inversion, each with a different emphasis on the training of surrogate models. The first approach relies exclusively on prior-based training, which we refer to as Full-Bandwidth-Prior-Training (FBPT). The second approach is based on the progressive enrichment of the training set with samples that are more representative of the posterior pdf, which we refer to as Progressively-Expanded-Posterior-Training (PEPT).

\subsection{Full-Bandwidth-Prior-Training (FBPT) methodology}
In recent work, \citet{meles2022bayesian} proposed a surrogate strategy for traveltime inversion based on PCA applied to both the input, permittivity distributions, and the output, traveltime data. It is natural to consider a similar approach for waveform inversion. However, the transition from traveltime data to waveform gathers is not straightforward in terms of output parameterization. To ensure effective output analysis, the FBPT strategy involves monochromatic passband filtering of the waveform data before performing PCA. Through this important preprocessing step, the FBPT strategy - though applied to waveform data -closely mirrors the approach outlined in \citet{meles2022bayesian}. We use \( \Pi \) to refer to input-related PCA concepts and \( \Lambda \) for output-related concepts. To enhance readability and facilitate comparison between the two schemes, we will also use 'F' and 'P' to represent the quantities associated with FBPT and PEPT, respectively.

\begin{table}[h]
    \centering
    \begin{minipage}{0.45\textwidth} 
        \centering
        \caption{FBPT symbols and quantities}
        \begin{tabular}{|c|c|}
        \hline
        \textbf{Symbol} & \textbf{Quantity} \\ 
        \hline
         \( I_F \) & Input training set \\  \hline
         \( N_F \) & Number of samples of \( I_F \) \\  \hline
         \( O_F \) & Output associated with \( I_F \) \\ \hline
        $\Pi_{F}$ & Number of input PCs \\ \hline
        $f_i$ & $i^{th}$ frequecny bin \\ \hline
        $\Omega_F$ & Set of frequency bins \\ \hline
        $\left| \Omega_F \right|$ & Cardinality of $\Omega_F$ \\ \hline   
        \( \hat{O}_F \) & Passbanded filtered \( O_F \) \\ \hline
        \( \Lambda_{\gamma} \) & Number of minigathers per frequency bin \\ \hline   
        $\gamma$ & Number of traces per minigather \\ \hline   
        $\widehat{\Lambda}_{F}$ & Number of PCs per minigather \\ \hline   
         $\Lambda_{F}$ & Number of PCs per frequency bin \\ \hline   
        $\Lambda_{FT}$ & Total number of PCs \\ \hline   
        $PCE_{F}$ & PCE model \\ \hline   
        $OBS$ & Observed traces to be inverted  \\ \hline   
        $OBS_{F}$ & Projection of OBS on $\Lambda_{FT}$ PCs  \\ \hline   
        $\mu_{F}$ & Proposal scale factor - $k^{th}$ iteration  \\ \hline   
        \end{tabular}
        \label{tab:FBPTSymbol}
    \end{minipage}
    \hspace{0.05\textwidth} 
    \begin{minipage}{0.45\textwidth}
        \centering
        \caption{PEPT symbols and quantities}
        \begin{tabular}{|c|c|}
        \hline
        \textbf{Symbol} & \textbf{Quantity} \\ 
        \hline
         \( k \) & PEPT Iteration number \\  \hline
         \( I_k \) & Input training set - $k^{th}$ iteration \\  \hline
         \( N_k \) & Number of samples of \( I_k \) \\  \hline
         \( O_k \) & Output associated with \( I_k \) \\ \hline
        $\Pi_{k}$ & Number of input PCs \\ \hline
        $f_i$ & $i^{th}$ frequecny bin \\ \hline
        $\Omega_k$ & Set of frequency bins - $k^{th}$ iteration \\ \hline
        $\left| \Omega_k \right|$ & Cardinality of $\Omega_k$ \\ \hline   
        \( \hat{O}_k \) & Passbanded filtered \( O_k \) \\ \hline
        \( \Lambda_{\gamma} \) & Number of minigathers per frequency bin \\ \hline   
        $\gamma$ & Number of traces per minigather \\ \hline   
        $\widehat{\Lambda}_{k}$ & Number of PCs per minigather \\ \hline   
        $\Lambda_{P}$ & Number of PCs per frequency bin \\ \hline   
        $\Lambda_{Pk}$ & Total number of PCs - $k^{th}$ iteration \\ \hline   
        $PCE_{k}$ & PCE model - $k^{th}$ iteration \\ \hline   
        $OBS_{k}$ & Projection of OBS on $\Lambda_{Pk}$ PCs  \\ \hline   
        $\mu_{k}$ & Proposal scale factor - $k^{th}$ iteration  \\ \hline   
        $P_{k}$ & Posterior - $k^{th}$ iteration \\ \hline   
        $\nu_{k}$ & Samples extracted from $P_k$  \\ \hline   
        $\rho_{p}$ & Samples kept from $P_p$  \\ \hline   
        \end{tabular}
        \label{tab:PEPTSymbol}
    \end{minipage}
\end{table}

The non-iterative FBPT process involves:

\begin{itemize}
    \item \textbf{Parameterization}: The first step is to define appropriate parameterizations for both the input and output domains. For the input, we consider  a training set \( I_F \) with \( N_F \) input samples, and parametrize it via its first  $\Pi_{F}$ PCs. For the output we collect a sample dataset $O_F$ of waveform data associated with $I_F$. We then consider frequency bins $f_i$ from a comprehensive set $\Omega_F$ capturing a relevant portion of the spectrum of the source term. 
    In a preprocessing step, the data \( O_F \) is passed through a passband filter for each \( f_i \in \Omega_F \). Such filtered data are denoted in the following as \( \hat{O}_F \).
    Sets of  $\widehat{\Lambda}_{F}$ PCs are derived for a number \( \Lambda_{\gamma} \) of time-domain minigathers, which are composed of subsets of \( \gamma \) adjacent common-source traces of \( \hat{O}_F \).
    These PCs are then used to parametrize the output. The data  \( O_F \) is thus transformed into the projection onto $\Lambda_{FT} =  \widehat{\Lambda}_{F} \times \Lambda_{\gamma} \times \left| \Omega_F \right|$ PCs 
    with $\left| \Omega_F \right|$ indicating the cardinality of $\Omega_F$. Each frequency bin results in ${\Lambda}_{F}=\widehat{\Lambda}_{F} \times \Lambda_{\gamma}$ output PCs. 
    \item \textbf{Model Training}: The second step is to train a surrogate model to map from input to output. We follow \citet{meles2022bayesian} and rely on PCE  modeling (see Appendix \ref{Appendix:PCE} for more details).
    The surrogate model, indicated here as $PCE_{F}$, estimate the projection on ${\Lambda}_{F}= \widehat{\Lambda}_{F} \times \Lambda_{\gamma}$ output PCs of each frequency bin of waveform data. Projection of a dataset $D$ on the chosen output PCs is indicated here as $D_{F}$. 
    \item \textbf{MCMC Inversion}: In the third step, $PCE_{F}$ is used as a surrogate model to perform MCMC inversion of the observed data $OBS$ projected on the considered output PCs, indicated as $OBS_{F}$, and estimate the corresponding posterior distribution $P_F$. Here, we use an Metropolis-Hastings algorithm with a Gaussian proposal distribution. The proposal scales are determined by a diagonal covariance matrix, where each diagonal element is proportional to the prior marginal variance of the corresponding input parameter, scaled by a factor indicated as $\mu_F$, determined through trial and error, aiming for an acceptance rate of around 20\%.
\citep{hastings1970monte, marelli2014uqlab}.
    \end{itemize}

A list of the most relevant symbols used in the FBPT algorithm can be found in Table \ref{tab:FBPTSymbol}.

In the FBPT scheme, a single surrogate model is used to predict the output of a waveform simulation across the prior pdf. Given the complexity of the input-output relationship, the training set should effectively represent the full range of the prior. It can be difficult to capture the variability of the prior using $N_F$ samples. In these cases, using $N_F$ samples from an \textit{inflated} prior can  be advantageous. This approach helps to avoid under-sampling in the peripheral regions of the prior. While it may affect the accuracy in well sampled areas, it ensures that the surrogate model is less prone tobias while still being sufficiently accurate across the entire prior. In this work, prior inflation is achieved by sampling from a Gaussian field with the same mean and correlation structure as the original prior, but with the standard deviations of each principal component increased by a factor $\alpha_F$.

\begin{algorithm}
\caption{Full-Bandwidth-Prior-Training (FBPT) scheme}\label{A:FBPT}
\begin{algorithmic}[1]
    \STATE Define a comprehensive set of frequency bins $\Omega_F$;
    \STATE For a training set \( I_F \) with \( N_F \) input samples from the prior pdf, possibly inflated by the factor \( \alpha_F \), a corresponding dataset \( O_F \) of waveform data is collected and passed through a passband filter for each frequency bin in \( \Omega_F \) to obtain \( \hat{O}_F \).
    \STATE Parametrize input via $\Pi_{F}$ PCs based on $I_F$;
    \STATE Parametrize \( \hat{O}_F \), represented by  $\Lambda_{\gamma}$  minigathers consisting of $\gamma$ adjacent receiver traces, via  $\Lambda_{FT} =  {\Lambda_{F}} \times \left| \Omega_F \right|$ PCs 
    with $\left| \Omega_F \right|$ indicating the cardinality of $\Omega_F$ and ${\Lambda}_{F}=\widehat{\Lambda}_{F} \times \Lambda_{\gamma}$, where  $\widehat{\Lambda}_{F}$ is the number of PCs per minigather. 
    \STATE Train $PCE_{F}$ using projections of $I_F, \hat{O}_F$ on $\Pi_{F}/{\Lambda}_{FT}$ input/output PCs. Using frequency depending output PCs 
    results in the projection of a generic dataset $D$  on a reduced dataset $D_F$;
    \STATE Use $PCE_{F}$ as a $\Pi_{F}$ dimensional  surrogate model to invert the $\Lambda_{F}$ dimensional dataset $OBS_{F}$, that is the projection of observed gather  $OBS$ and estimate posterior $P_F$.
\end{algorithmic}
\end{algorithm}

Note that each output component - corresponding to a specific frequency and a specific minigather - requires its own surrogate model. The term \textit{PCE model} thus refers to the collection of all these individual output models.

\subsection{Progressively-Expanded-Posterior-Training (PEPT) methodology}

As an alternative to FBPT, the PEPT scheme adopts a sequential model refinement approach. It starts by constructing a PCE surrogate model that inverts only the low-frequency content of the data to constrain low-wavelength components in the input space. This surrogate model is then used to perform an initial MCMC inversion. The resulting posterior samples are employed to refine the surrogate model, assuming that the target posterior lies within the initial posterior. This assumption is reasonable because PCE is unbiased over its training set, only low-frequency data are used and the inclusion of model errors in the likelihood function leads to a more dispersed posterior in the first inversion due to the relatively large model errors. Using samples from this more constrained region of the input domain, we then retrain the surrogate model with an expanded frequency bandwidth.
Multiple iterations are carried out to achieve a final surrogate model that adequatelysimulate each data component of interest in the posterior region. Once the desired level of accuracy is attained, the resulting final surrogate model is employed to conduct a final MCMC inversion to approximate the posterior of interest. 
This iterative PEPT strategy demands additional considerations.

\begin{itemize}
    \item \textbf{Parameterization}: As for FBPT, the first step is to define an appropriate parameterization for the input and output domains. To ensure a fair comparison, the same  input parameterization will be applied to both the FBPT and PEPT strategies and the same number of output PCs, given a frequency bin,  will be used for both methods.  
\end{itemize}

After this preliminary step, a process follows that is repeated \textit{K} times:

\begin{enumerate}
    
    \item \textbf{Subset Selection}: In each iteration $k, 1 \leq k \leq K$, different subsets of the input and output domains are utilized. This is achieved by progressively increasing the number of input principal components ($\Pi_{P_1}, \dots, \Pi_{P_K}$) and frequency bins ($\Omega_1, \dots, \Omega_K$) on which the data is projected in the preprocessing step to obtain  \( \hat{O_k} \). As in FBPT, the total number of minigathers ($\Lambda_{\gamma}$) and the corresponding output principal components ($\Lambda_{P}$) can also be adjusted. In this manuscript, to ensure a fair comparison between the two methods, we apply the same choices for both across all iterations of PEPT; however, this is by no means a requirement. Given the number of minigathers ($\Lambda_{\gamma}$) and the corresponding PCs  ($\widehat{\Lambda}_{P}$), the total number of output PCs per frequency bin is ${\Lambda}_{P}=\widehat{\Lambda}_{P} \times \Lambda_{\gamma}$, and the total number of output PCs $\Lambda_{PT_k} = {\Lambda_{P}} \times \left| \Omega_k \right|
$, where $\left| \Omega_k \right|$ indicates the cardinality of $\Omega_k$, i.e. the number of frequency bins.
    In terms of dimensionality, the input and output domains used for inversion in PBET will fully align with those in FBPT during the final iteration, where $\Pi_{P_n} = \Pi_{F}$ and $\Omega_{n} = \Omega_F$ and therefore $\Lambda_{FT}=\Lambda_{PT_K}$. At this final stage, both FBPT and PBET utilize the same set of input PCs. However, despite having the same dimensionality, the output PCs will differ between the two methods, as for FBPT they are defined based on samples drawn from the prior, while for the PEPT they are based on samples drawn from progressively refined posteriors. Updating the output PCs is key to ensure optimal representation of the relevant data variability in high-posterior regions.
    \item \textbf{Model Training}: In each iteration $k, 1 \leq k \leq K$
, a surrogate model is trained to map the relevant input and output subdomains. The training set needs to be large enough to achieve sufficient accuracy while not too large to keep computational costs low. Initially, samples are drawn from $I_1$, $\hat{O}_1$.
    In subsequent iterations, training focuses primarily on the progressively refined, narrower subdomains explored in the previous MCMC inversion detailed below. The surrogate model, indicated here as $PCE_{k}$, estimate the projection of a dataset on ${\Lambda}_{P}= \widehat{\Lambda}_{P} \times \Lambda_{\gamma}$ output PCs per each considered frequency bin of waveform data.
    \item \textbf{MCMC Inversion}:
$PCE_{k}$ is used as a surrogate model to perform MCMC inversion of the observad data $OBS$ projected on the considered output PCs, indicated as $OBS_{k}$, and estimate the corresponding posterior distribution $P_k$.
To enhance the efficiency and convergence properties of the MCMC algorithm, we adjust the proposal distribution at each PBET iteration based on the correlations among the different input components, as inferred from the previous posterior sample. 
Unlike the adaptive scheme by \citet{haario2001adaptive}, where the proposal distribution is dynamically adjusted as the chains progress, we modify the proposal distribution only during the PEPT iterations, relying instead on a standard MH scheme. The proposal scales are determined by a diagonal covariance matrix, where each diagonal element is proportional to the prior marginal variance of the corresponding input parameter, scaled by a factor indicated as $\mu_{Pk}$ determined through trial and error, aiming for an acceptance rate of around 20\%. Once established in preliminary tests, this scaling is applied to each inversion run discussed here.
Given this adaptation of the proposal scheme, the covariance structure is shared between the PEPT and FBPT schemes only at the initial iteration of the PEPT.
At the final iteration of PEPT, the same data as in FBPT are inverted, but the tuning of their respective MCMC schemes may differ significantly. 
To minimize burn-in, it is possible to initialize the chains using draws from MCMC runs of previous iterations. For instance, the first iteration produces a posterior for the first $\Pi_{P1}$ PCs. The final values from this step can serve as the starting point for the first iteration, while the remaining $\Pi_{P2}$-$\Pi_{P1}$ PCs starting values are drawn from the prior.
\item \textbf{Training Set Enrichment}: If $k \leq K$, it is necessary to enhance the training set by drawing additional samples from the MCMC posterior to improve the model’s ability to describe broader frequency bandwidths of the data. 
This involves collecting $\nu_k$ samples from $P_k$, adding components from the \textit{null space}, defined as the prior subdomain not considered in the modeling/inversion process \citep{meles2022bayesian}, and transforming them back into the physical distribution domain. These new permittivity distributions are then included  in $I_{k+1}$, potentially alongside samples  samples $\rho_p, p \leq k$  from previous iterations and used to compute $O_{k+1}$. Note that including samples from earlier iterations — associated with broader posteriors may reduce PCE accuracy in the regions of primary interest. Therefore, their use must be carefully evaluated.
\end{enumerate}

A list of the most relevant symbols used in the PEPT algorithm, along with their meanings, can be found in Table \ref{tab:PEPTSymbol}.

As in FBPT, an inflation factor $\alpha_P$ can be applied during training to mitigate excessive inaccuracies in surrogate modeling performance.

\begin{algorithm}
\caption{Progressively-Expanded-Posterior-Training (PEPT) scheme}\label{A:PEPT}
\begin{algorithmic}[1]
    \STATE Define expanding sets of frequency bins ($\Omega_1$, ... $\Omega_{n}$);
    \STATE For a sample $I_1$ with \( N_1 \) input samples (typically with $ N_1 \ll N_F$ ) from the  prior pdf, possibly inflated by a factor $\alpha_P$, collect a dataset $O_1$ of waveform data and and passed through a passband filter for each frequency bin in $\Omega_1$ to get  \( \hat{O_1} \).;
    \STATE Parametrize input (via $\Pi_{P_1}$ PCs based on $I_1$);
    \STATE Parametrize  \( \hat{O_1} \), represented by  $\Lambda_{\gamma}$  minigathers consisting of $\gamma$ adjacent receiver traces, via  $\Lambda_{P1} =  {\Lambda_{P}} \times \left| \Omega_1 \right|$ PCs 
    with $\left| \Omega_1 \right|$ indicating the cardinality of $\Omega_1$ and ${\Lambda}_{P}=\widehat{\Lambda}_{P} \times \Lambda_{\gamma}$, where  $\widehat{\Lambda}_{P}$ is the number of PCs per minigather.    
    \FOR {$k = 1$ \TO $n$}
    \STATE Train $PCE_{k}$ using projections of $I_k,\hat{O}_{k}$  on ${\Pi}_{Pk}$, ${\Lambda}_{Pk}$ input/output PCs. Using frequency depending output PCs 
    results in the projection of a generic dataset $D$ on a reduced dataset $D_k$;
    \STATE Use $PCE_{k}$ to invert $OBS_{k}$, that is, the projection of observed gather  $OBS$ and estimate posterior $P_k$;
    \STATE Collect $\nu_k$ samples from $P_k$, add elements from the \textit{null space} and transform back into the physical distribution domain and include them in $I_{k+1}$. 
    \STATE Based on empirical accuracy and convergence considerations, samples $\rho_p$ from $I_{p}, p \leq k $ can also be included in $I_{k+1}$;
    \STATE Given $I_{k+1}$ compute with the reference forward solver simulations corresponding output $O_{k+1}$  and pass it through a passband filter for each frequency bin in $\Omega_{k+1}$ to get \( \hat{O}_{k+1} \);
    \STATE Parametrize input (via $\Pi_{P{k+1}}$ PCs based on $I_{k+1}$);
    \STATE Parametrize \( \hat{O}_{k+1} \), represented by  $\Lambda_{\gamma}$  minigathers consisting of $\gamma$ adjacent receiver traces, via  $\Lambda_{P{k+1}} =  {\Lambda_{P}} \times \left| \Omega_{k+1} \right|$ PCs 
    with $\left| \Omega_{k+1} \right|$ indicating the cardinality of $\Omega_{k+1}$ and ${\Lambda}_{P}=\widehat{\Lambda}_{P} \times \Lambda_{\gamma}$, where  $\widehat{\Lambda}_{P}$ is the number of PCs per minigather.  
    \ENDFOR
\end{algorithmic}
\end{algorithm}

Due to the orthogonality of PCA and Fourier decomposition, both the FBPT and the PEPT approaches ensure that orthonormality between corresponding basis vectors is maintained when considering mutually separated bandwidths. Consequently, the data error and the likelihood structure in the original gather domain are preserved in these specific output quantities (see Appendix in \citet{meles2022bayesian}). Note that reducing the number of PCs in the input space alters the covariance matrix in the likelihood term, whereas restricting the number of PCs in the output domain does not affect the likelihood structure.     


\begin{figure}
\centering
\includegraphics[width=1\textwidth]{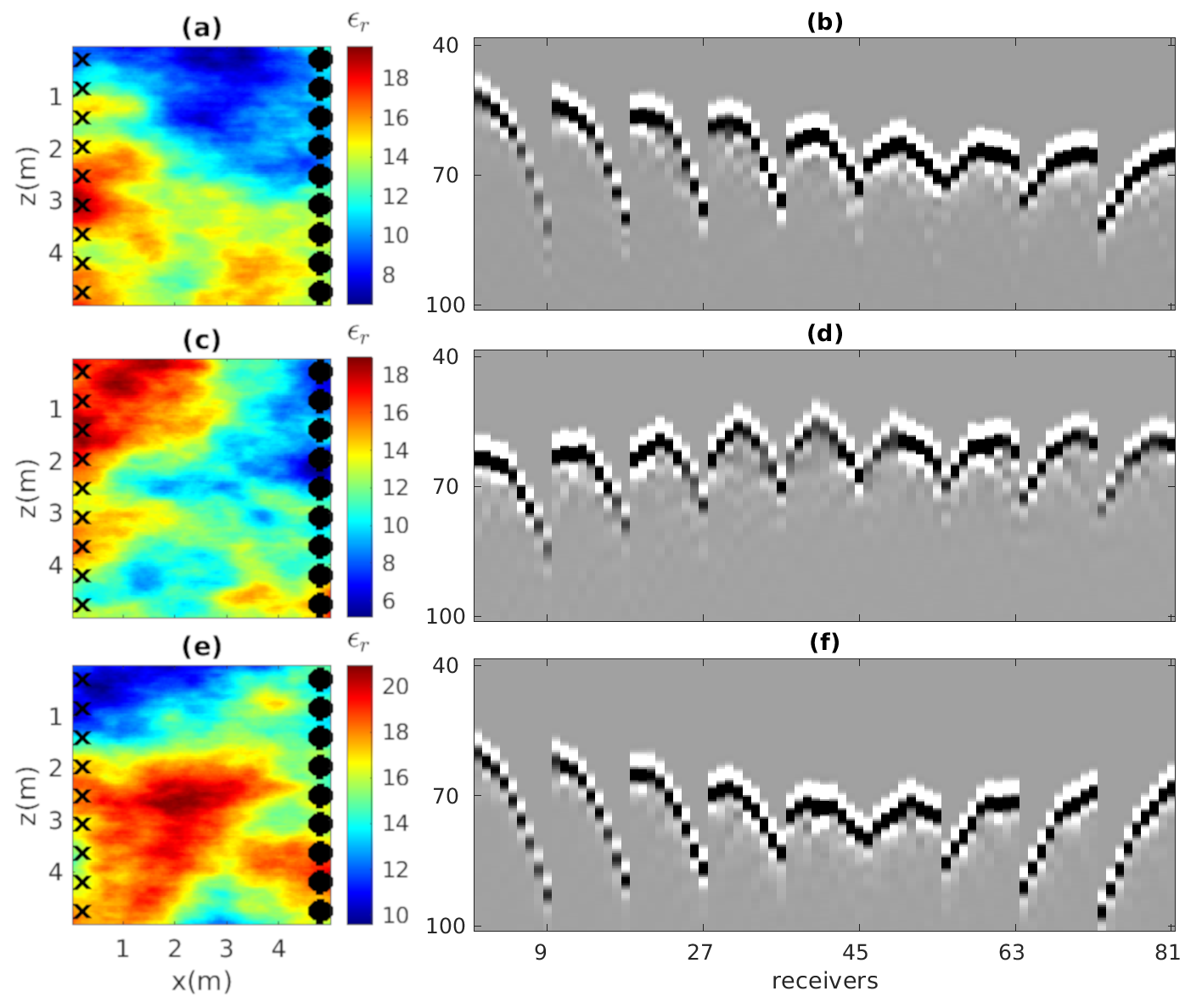}
\caption{\label{FIG_001}}
(a, c, e) Selected realizations from the generative model and (b, d, f) the respective waveforms at receiver locations. Source and receiver locations are marked (a, c, e) with crosses and circles, respectively. 
\end{figure}

\section{Application to GPR crosshole tomography}

We apply the proposed FBPT and PEPT methodologies to a GPR crosshole tomography problem \citep{ernst2007full,Kuroda2007,meles2010new}. We consider for simplicity variations in relative permittivity, while assuming a constant and known electrical conductivity. We use the recording configuration displayed in Fig. \ref{FIG_001}(a) with 9 evenly spaced sources and receivers located in two vertically-oriented boreholes. The distance between the boreholes is 4.6 m, while the spacing between sources/receivers is 0.56 m. Each source is characterized by a $100$ MHz Blackman–Harris pulse that is initiated separately and the propagating signals are recorded at all receivers. We employ a 2D FDTD solver to simulate noise-free propagation in the transverse-electrical (TE) mode \citep{irving2006numerical} with perfectly matched layers. We rely on a generative model representing prior knowledge to create a set of relative permittivity distributions $\epsilon_r$ typical of a near-surface geophysical setting \citep{hunziker2017inference}. We consider $5$ m wide, $5$ m deep random multi-Gaussian field realizations created using a pre-defined covariance structure based on the 2-D Matérn geostatistical model \citep{dietrich1997fast,laloy2015probabilistic}. The models are generated using six parameters: mean ($14$), standard 
deviation of $\epsilon_r$ ($3$), anisotropy angle ($90$), anisotropy ratio ($0.5$), integral scale of the major axis (10 m) and a shape parameter ($1$). 
Figures \ref{FIG_001}(a, c, e) display representative samples of $\epsilon_r$-distributions, together with the corresponding gathers in Figs. \ref{FIG_001}(b, d, e). 
Given the permittivity range and the frequency content of the source wavelet, a spatial grid with $dx=dz=4$ cm and a time stepping dt of $0.16$ \text{ns} is used to avoid numerical dispersion and instability. Each generated permittivity field corresponds to a $125\times125$ dimensional space. 
\label{Generative_Model}

\begin{figure}
\centering
\includegraphics[width=1\textwidth]{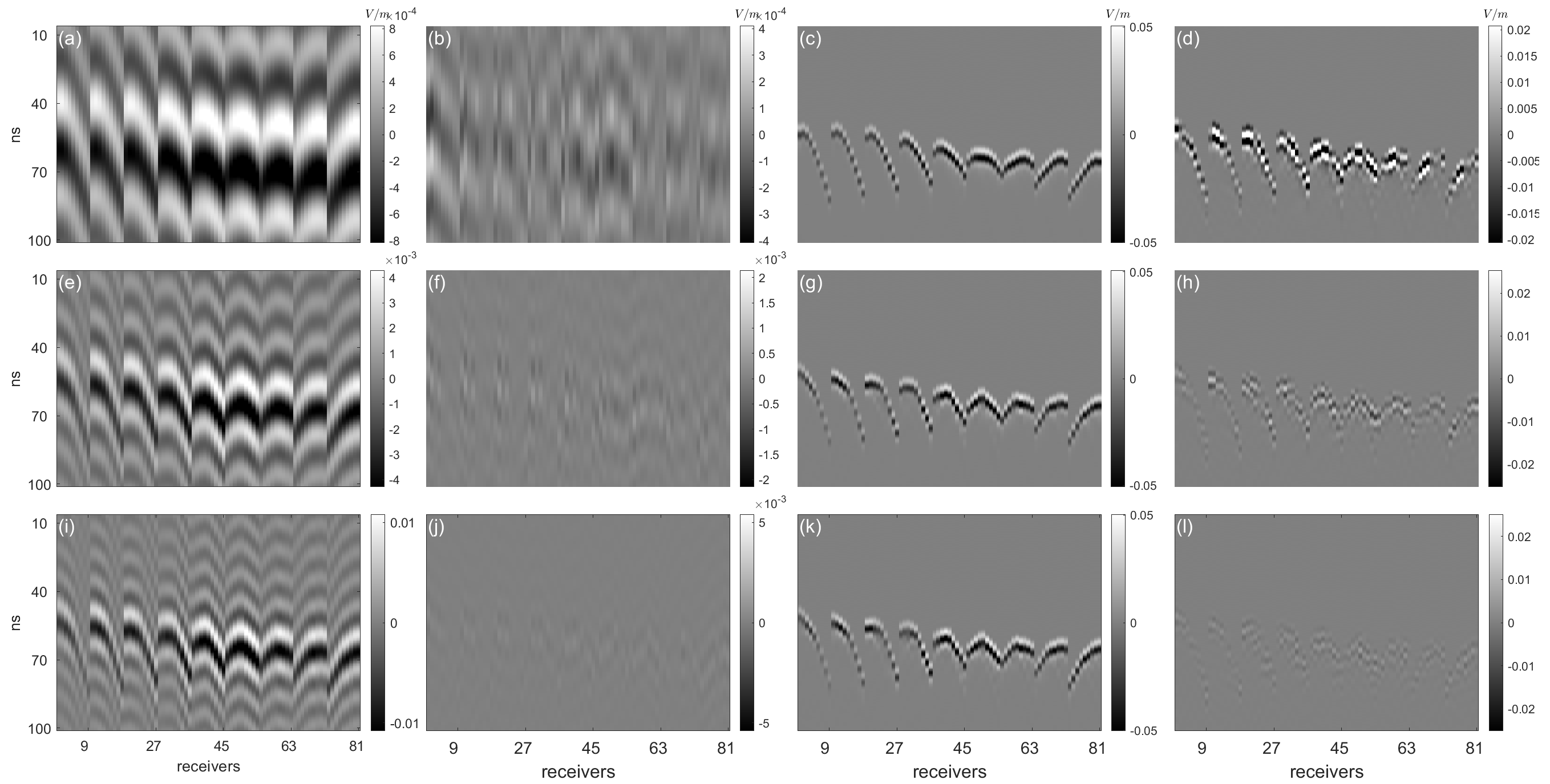}
\caption{\label{FIG_003} (a) Low-frequency (0-18 MHz) passband-filtered forward solution when the input is the projection of the relative permittivity distribution in Fig. 1(a) onto the first five PCs.
(b) For the same frequency bandwidth as (a), the difference with respect to the data corresponding to the forward solution for the reference relative permittivity distribution in Fig. 1(a).
(c) and (d) As in (a) and (b), but considering the full bandwidth of the data.
(e)-(h) As in (a)-(d), but using 35 PCs and the 0-45 MHz bandwidth.
(i)-(l) As in (a)-(d), but using 100 PCs and the 0-72 MHz bandwidth.}
\end{figure}

\label{parametrization}
Following \citet{meles2022bayesian}, we use the generative model to create a total of 1000 random realizations from the prior and learn a prior-knowledge-based parametrization of the input domain (i.e., $\epsilon_r$ distributions) in terms of principal components, with the first ones linked to low-wavenumber features, and higher modes revealing small-scale details. In terms of wave propagation, the low-frequency content of the data is mainly sensitive to the first principal components. To also capture the high-frequency content in the output, it is necessary to expand and consider the higher modes of the principal components. This is demonstrated in Fig. \ref{FIG_003} using the reference $\epsilon_r$ distribution in Fig. \ref{FIG_001}(e) and its truncations on 5, 35 and 100 PCs, and the corresponding data computed via FDTD modeling for low (0-18 MHz), medium (0-45 MHz), high (0-72 MHz) and full frequencies (0-300 MHz) bandwidths. The ability of the truncated input to capture the reference data set depends on the frequency content considered, with 5 PCs being able to only adequately recover low frequencies of the data (Fig. \ref{FIG_003}(a-d)). As we expand the PCA basis elements up to 35, the residual  reduces even when considering an expanded frequency bandwidth (Fig. \ref{FIG_003}(e-h)). Projecting the input on 100 PC components results in close agreement across the entire bandwidth of the data (Fig. \ref{FIG_003}(i-l)).

The standard output in full waveform applications consists of all values in the inverted gathers. For the problem considered here, with gathers ranging from 0 to 110 ns and a time sampling interval of \(0.32\ \text{ns}\), this results in a total of \(9 \times 9 \times 344 = 27864\) values.

For the PCE implementation, the input parametrization is guided by the effective dimensionality analysis discussed above. Based on our analysis, we use the decomposition of the relative permittivity distribution on the first 100 input PCs ($ \Pi_F = 100$ in Algorithm \ref{A:FBPT}).

To reduce the output dimension and build an effective surrogate, we apply PCA alongside frequency filtering of the data. Key parameter choices were made by a performance analysis. These choices include the data subdomains and the frequency bandwidths used for defining the output principal components, and the number of principal components considered for modeling and inversion. The selection process involved evaluating the method’s behavior across various configurations to optimize accuracy and computational efficiency.

We use mutually disjoint passband filters and consider each frequency bin individually. We then apply PCA decomposition to minigathers, each consisting of 3 ($\gamma$=3  in algorithm \ref{A:FBPT}) adjacent receiver traces for each filtered dataset. In total, 27 such minigathers ($\Lambda_{\gamma}$=27 in algorithm \ref{A:FBPT}) compose the entire dataset for one frequency bin, which is trained independently of the other frequencies. For each frequency and minigather, we use two output PCs ($\widehat{\Lambda}_{F}$=2 in algorithm \ref{A:FBPT}) per minigather  per frequency bin, which allowed us to capture a considerable amount of the variance in the data while retaining high accuracy of the surrogate modeling. The first two components capture, on average, about 60\% of the total variance across the various frequency bins and minigathers considered here.



\begin{figure}
\centering
\includegraphics[width=1\textwidth]{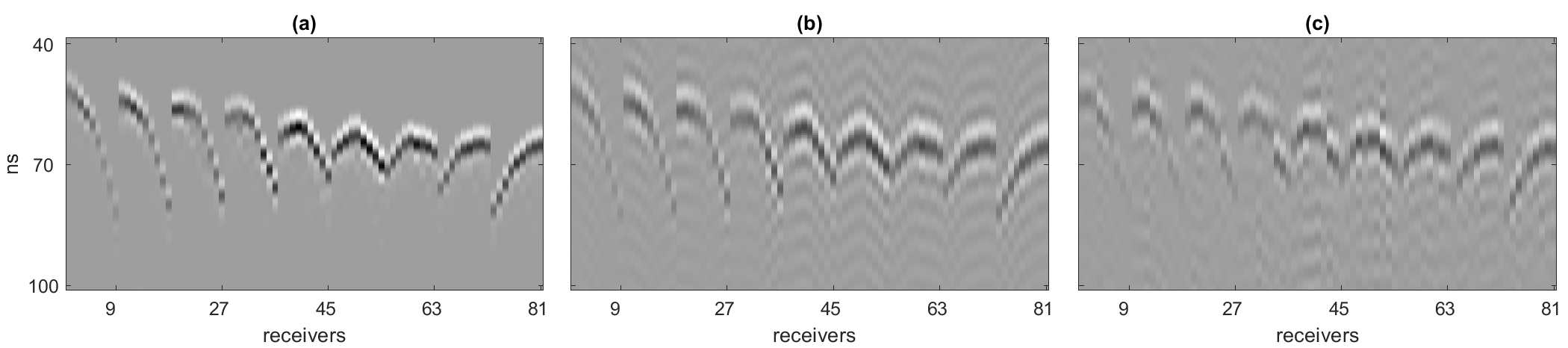}
\caption{\label{FIG_FBPT} (a) Full-bandwidth response for the relative permittivity distribution shown in Fig. \ref{FIG_001}(a). (b) Passband-filtered response of (a) within the 0–144 MHz bandwidth.
(c) Projection of (b) onto the 864 output PCs considered in the FBPT scheme.} 
\end{figure}

For the FBPT scheme, a total of 16 equally spaced frequency bins, represented by the parameter $\Omega_F$ in Algorithm~\ref{A:FBPT}, are used. These bins correspond to a minimum and maximum frequency of 9 MHz and 144 MHz, respectively. The parameter $\Omega_F$ in Algorithm~\ref{A:FBPT} is defined as  $\Omega_F = \{9, 18, 27, \dots, 144\}$.  
 This corresponds to a total of 864 ($\Lambda_{FT}$=864 in algorithm \ref{A:FBPT}) output data (2 PCs $\times$ 27 minigathers $\times$ 16 bins) to be inverted. A visual comparison of the full-bandwidth, passband-filtered, and projected datasets used in the FBPT is shown in Fig. \ref{FIG_FBPT}.

 The training is performed on a total of 2,900  samples from a broad distribution centered on the prior by a factor $\alpha_F$=2. Expanding the input space leads to a decrease in \textit{overall} accuracy across the effective prior pdf when the number of training sets is fixed. However, we find that the enlargement avoids the introduction of bias on specific target distributions. The enlargement factor $\alpha_F$  was chosen based on trial-and-error analysis. A summary of the parameters used in the FBPT experiment can be found in Table \ref{tab:FBID}.

The parameters chosen for the PEPT experiment, most notably the number of PCs in the input and output, and the inverted frequencies, were determined through a trial-and-error approach, guided by heuristic considerations and empirical observations.

For the PEPT strategy, we begin with a $1^{\text{st}}$ iteration based on draws from the same inflated prior used for the FBPT scheme, using 900 samples and the first three frequency bins, denoted as $\Omega_1$ in Algorithm \ref{A:PEPT}, corresponding to 9, 18, and 27 MHz. For each frequency bin, we consider the first two PCs of the output domain for each minigather consisting of three adjacent receivers ($\Lambda_{\gamma}$=27 and $\widehat{\Lambda}_{P}$=2) in Algorithm \ref{A:PEPT}. This results in a total of 162 waveform attributes used in the inversion ($\Lambda_{P1}$=162 in Algorithm \ref{A:PEPT}). Since we are inverting low-frequency data in this initial stage, we consider only the first $15$ PCs of the input domain ($\Pi_{P1}$=15 in Algorithm \ref{A:PEPT}), as the remaining components have minimal impact on the response and are treated as part of the 'null space'. We then train a PCE model mimicking the relationship between the first $15$ input domain PCs and each of the analysed $162$ output attributes. 
From the posterior obtained in this first iteration, we extract $\nu_1=500$ samples that are used when retraining the PCE surrogate, with $\rho_1=0$ (see Algorithm \ref{A:PEPT}). This initial iteration of the algorithm employs an inflated  prior-based PCE model. Afterward, three additional iterations are performed, each leveraging the results of the previous inversion to enhance the training set in the region of significant posterior probability. For the second and third iteration, $\nu_{2,3}=500$ and $\rho_{2,3}=0$. For the fourth and final iteration, we used $\nu_{4}=500$ and $\rho_{3}=100$. These iterations focus on model expansion and refinement, similar to the first PEPT iteration but with a posterior-based PCE model, and with adjustments to input/output dimensions and MCMC parameters. While each iteration refines the posterior-based PCE model based on prior inversion outcomes, all other parameters, including input/output dimensions and MCMC settings, are predetermined by the iteration number, independent of the specific dataset. The entire inversion process is adaptive yet fully automated. A summary of the parameters used in each PEPT experiment iteration can be found in Table \ref{tab:PEPT}.
\begin{table}[ht]
\centering
\caption{Parameters used in the FBPT algorithm.}
\begin{tabular}{@{}lcccccr@{}}
    \toprule
    \textbf{Iteration} & $\Pi_{F}$ & $\Lambda_{FT}$   & \textbf{\#Chains} & \textbf{MCMC Steps} & \textbf{Bandwidth (bins)} & \textbf{Training Set (size)} \\
    \midrule
    1 & 100 & 864 & 10 &30000 & $\Omega = \{1,2,..,16 \}$ & $I_F/O_F (2900)$ \\
    \bottomrule
\end{tabular}
\label{tab:FBID}
\end{table}

\begin{table}[ht]
\centering
\caption{Parameters used for the PEPT algorithm.}
\begin{tabular}{@{}lcccccr@{}}
    \toprule
    \textbf{Iteration} & $\Pi_{k}$ & $\Lambda_{Pk}$  & \textbf{\#Chains} & \textbf{MCMC Steps} & \textbf{Bandwidth (bins)} & \textbf{Training Set (size)}  \\
    \midrule
    1 & 15 & 162 & 10 & 7500 & $\Omega_1 = \{1,2,3 \}$ & $I_1/O_1 (900)$ \\
    2 & 50 & 324 & 10 & 25000 & $\Omega_2 = \{1,2,..,6 \}$ & $I_2/O_2 (500)$ \\
    3 & 80 & 486 & 10 & 80000 & $\Omega_3 = \{1,2,..,9 \}$ & $I_3/O_3 (500)$ \\
    4 & 100 & 864 & 10 & 100000 & $\Omega_4 = \{1,2,..,16 \}$ & $I_3/O_3 (100)$ $\cup$ $I_4/O_4 (500)$ \\
    \bottomrule
\end{tabular}
\label{tab:PEPT}
\end{table}

\begin{figure}
\centering
\includegraphics[width=1\textwidth]{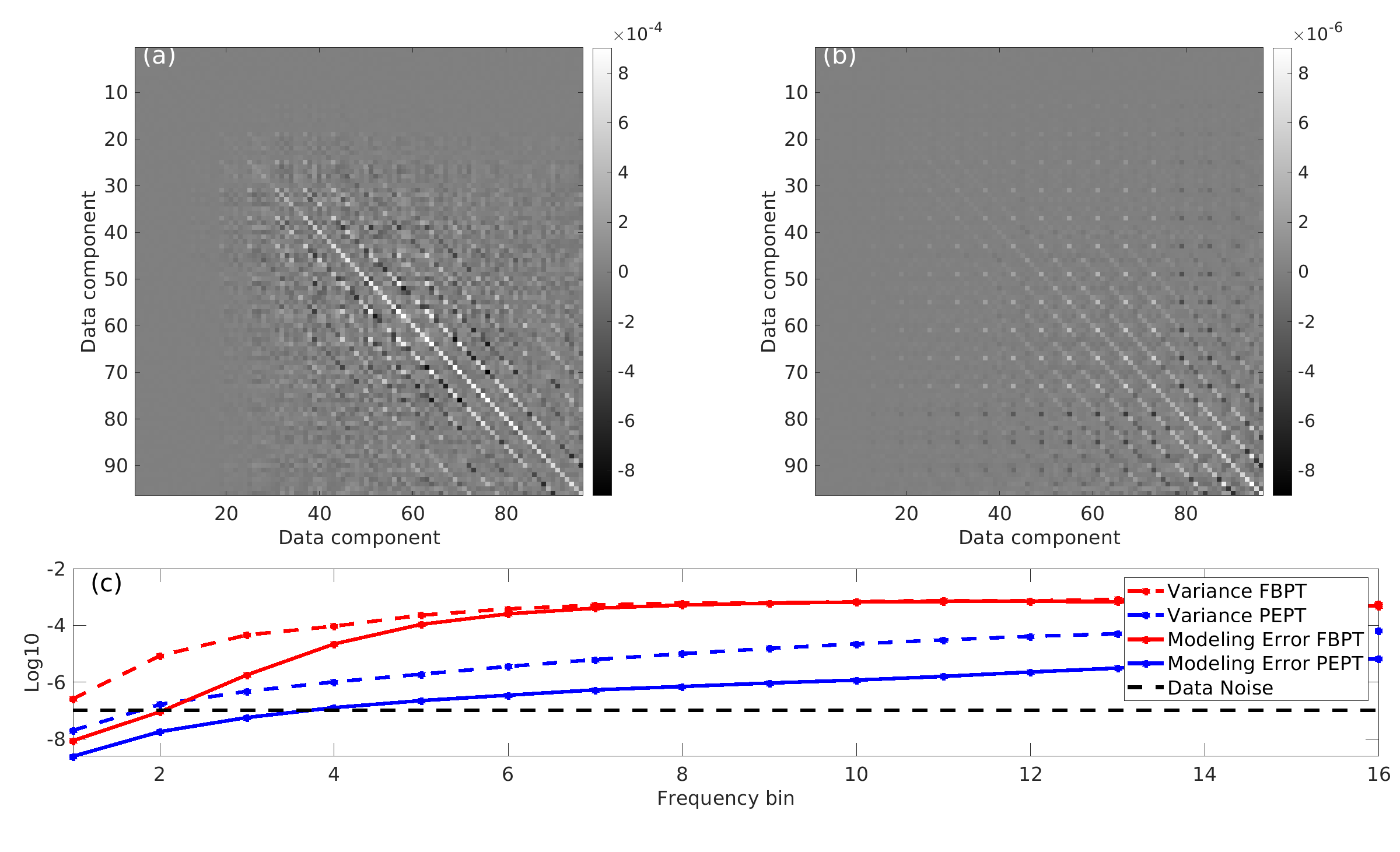}
\caption{\label{FIG_009} (a) Modeling error matrix for the FBPT scheme. For readability, only 96 out of 864 output values are shown. (b) Similar to (a), but for the fourth and final iteration of the PEPT scheme. (c) Mean values of the diagonal elements of the modeling error terms in (a) and (b) (red and blue solid lines, respectively), variance of the corresponding training sets (red and blue dashed lines, respectively), and variance of the noise terms (black dashed line) as a function of the frequency bin.}
\end{figure}

We evaluate the FBPT and PEPT schemes considering waveform data generated by the $\epsilon_r$ distribution in Fig. \ref{FIG_001}(a) and compare with MCMC inversion of the corresponding travel-times using an eikonal solver \citep{hansen2013sippi} using the UQLab Matlab package \citep{UQdoc_14_113}.

Uncorrelated Gaussian noise with a standard deviation equal to 2\% of the maximum amplitude of the gather is added to the synthetic FDTD data prior to PCA decomposition and inversion. The travel times are contaminated with uncorrelated Gaussian noise with a standard deviation of 0.5 ns.

Similarly to \citet{meles2022bayesian}, 
we account for the modeling and truncation error in the covariance operator \citep{hansen2014accounting,madsen2018estimation}.
While in \citet{meles2022bayesian}, the use of the modeling error term only slightly improved the inversion results, here it is absolutely crucial. This can be appreciated by analyzing the magnitude of the modeling error with respect to the noise level for the different frequency bins. Figures \ref{FIG_009}(a) and (b) show the modeling error terms for the FBPT and the final iteration of the PEPT schemes (for readability purposes, only one out of nine output components is shown). Figure \ref{FIG_009}(c) further shows, as a function of the frequency bin, the mean value of the diagonal terms of the modeling error covariance matrices (red and blue solid lines), the variance of the output of the corresponding training sets (red and blue dashed lines), and the noise level (black dashed line). These figures indicate that the modeling error associated with FBPT is considerably smaller than the noise level only for the first four frequency bins and that it reaches the variance of the training set around the sixth bin. In contrast, the modeling error associated with the last iteration of the PEPT scheme, despite being larger than the noise level for most of the considered frequency bins, is consistently an order of magnitude smaller than the variance of the corresponding training set. The important differences in the magnitudes of the noise and modeling error terms highlight the importance of including the latter in the likelihood to ensure reliable inversion.

\begin{figure}
\centering
\includegraphics[width=1\textwidth]{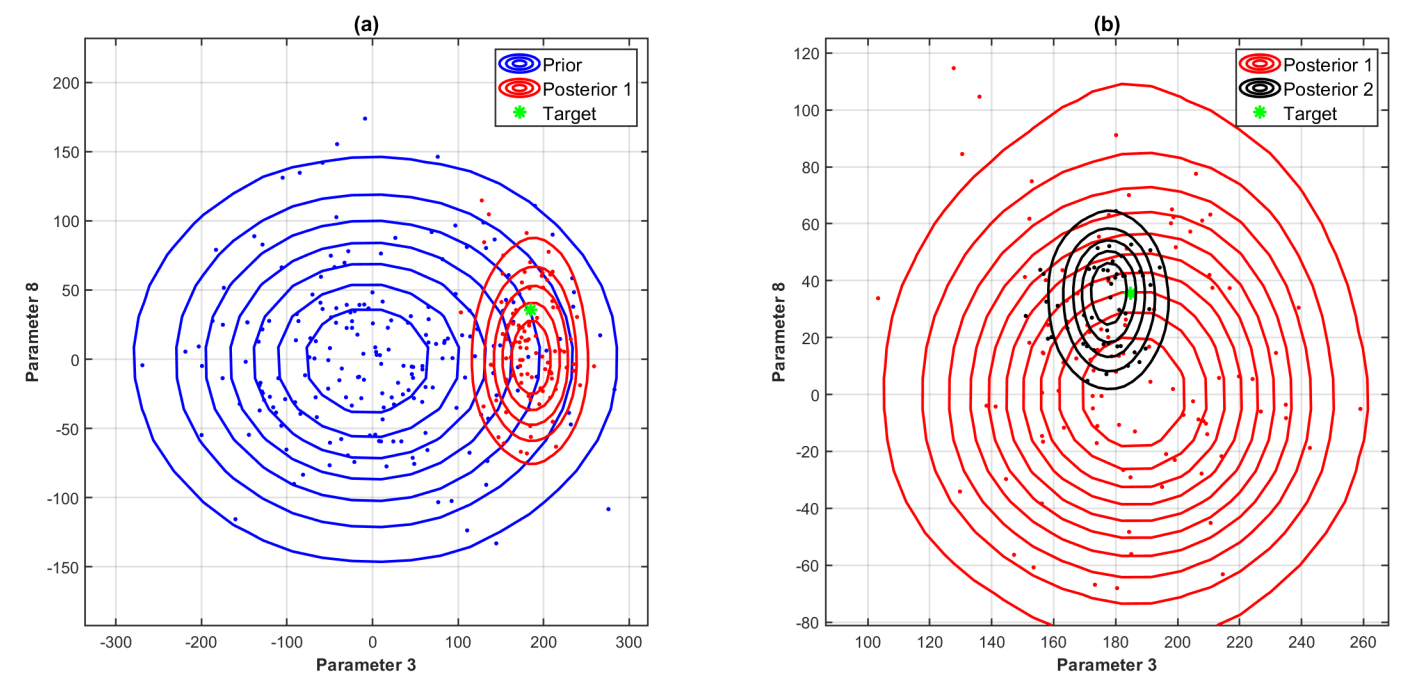}
\caption{\label{FIG_005} For the PEPT scheme, joint pdfs between the 3\textsuperscript{rd} and 8\textsuperscript{th} principal components for the prior (blue isosurfaces), 1\textsuperscript{st} (red isosurfaces), and 2\textsuperscript{nd} (black isosurfaces) posteriors. The green dot corresponds to the true values of the target input distribution. Blue and red dots correspond to training elements from the prior and posterior of the 0\textsuperscript{th} PEPT iteration. The black squares in (a) and (b) cover the same areas of the parameter space.}
\end{figure}

When analyzing Fig. \ref{FIG_009}(c), it is important to recognize that the curves correspond to different validation sets: the FBPT scheme is validated over the prior inflated by the factor $\alpha_F$=2, while the PEPT scheme focuses on a much narrower region centered around the posterior distribution. The substantial reduction in \textit{relative} modeling error observed with the PBET scheme arises from training on samples obtained through inversions of progressively expanded bandwidths. This strategy concentrates on a progressively smaller portion of the prior, ensuring convergence near the target and maximizing accuracy in the subsequent development of surrogate models in the most critical regions of the input space. This can be seen in Fig. \ref{FIG_005} for two arbitrarily chosen input parameters (whose exact values are indicated by the green dot), the progressive focusing around the target of posteriors as a function of the PEPT iteration is illustrated. For the 1\textsuperscript{st} iteration, the target lies in a peripheral zone of the prior on which the training takes place (blue isosurfaces in Fig. \ref{FIG_005}(a)). As such, we can expect somewhat low accuracy in the corresponding surrogate modeling. Still, inverting the low-frequency content of the data only, as prescribed by the PEPT scheme, allows focusing around the true parameter value (red isosurfaces in Fig. \ref{FIG_005}(a-b)). Further training leads to improved accuracy, which in turn results in an even more focused posterior as provided by the 2\textsuperscript{nd} iteration of the PEPT scheme (green isosurfaces in Fig. \ref{FIG_005}(b)). This trend continues throughout the PEPT scheme, ultimately leading to the reduced modeling error term discussed above and correspondingly to accurate inference.

\begin{figure}
\centering
\includegraphics[width=1\textwidth]{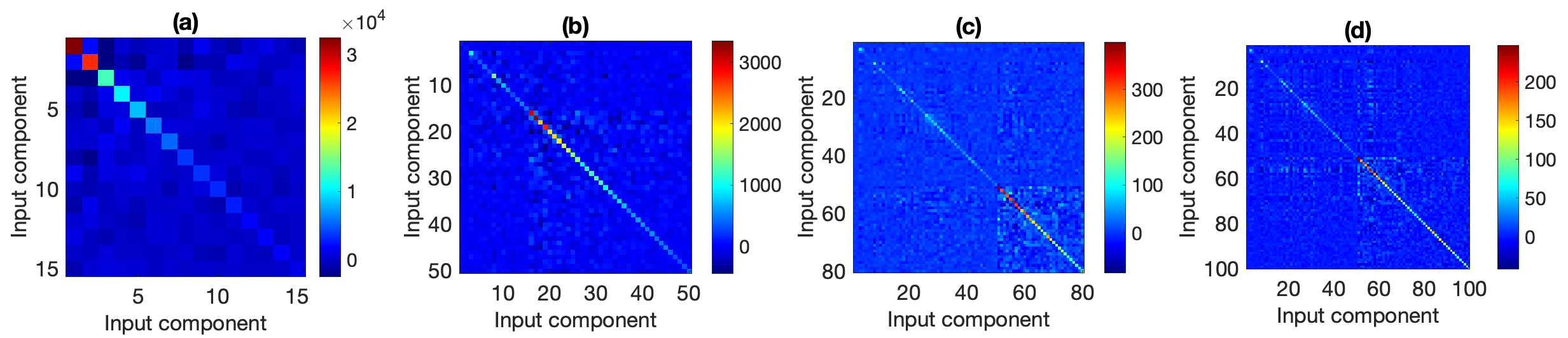}
\caption{\label{FIG_010} (a) Covariance matrix based on the validation set used in the first iteration of the PEPT scheme. (b)-(d) For the inversion associated with the permittivity distribution in Fig. \ref{FIG_001}(a), covariance matrices of the MCMC model proposals for the second to fourth PEPT iterations, showing a strong decrease in magnitude after each surrogate model update. 
These matrices highlight the significant differences between the subsequent PEPT matrices and what would have been used if the covariance matrices were not updated between the different iterations of the PEPT algorithm.}
\end{figure}

For the FBPT scheme, we use a MH algorithm based on a Gaussian proposal distribution scaled with respect to the prior distribution. 
For the PEPT $1^{st}$ iteration, we follow the same scheme as for the FBPT in terms of proposal distribution \label{Proposal}. For the remaining iterations, the correlations among the input components are inferred using the previous iteration  and used to build corresponding covariance proposal matrices. Examples of such matrices are shown in Fig. \ref{FIG_010}(b-d). While all are diagonally dominant, the matrices exhibit noticeable variability along the diagonal and in the off-diagonal elements. In particular, capturing the relative variations along the diagonal is crucial to enable adequate posterior exploration in the later inversion stages. As the iterations progress, the first PCs in the posterior becomes very well defined, which implies that maintaining the same scales as for FBPT would lead to very low acceptance rates. 
At each iteration the number of inverted parameters expands, the size of the matrices increases with each iteration, ranging from 15 to 100, and the magnitudes decrease.

The inversion results are assessed by comparing the posterior mean and Maximum A Posteriori (MAP) realization, including elements from the 'null space', and computing the Root Mean Square Error (RMSE), structural similarity (SSIM) and log score maps  \citep{gneiting2007strictly,friedli2022lithological}. The log score evaluates the accuracy of probabilistic predictions by quantifying how well a predicted posterior probability distribution matches the true values.
It refers to the negative logarithm of the inferred posterior PDF evaluated at the true value. This involves using kernel density estimation (KDE) on posterior samples to estimate the PDF, and then calculating the log of the density at the true value across the domain to generate the log score maps.
Mathematically, it is expressed as:
\[ \log S(\hat{P}, \vx_{\text{true}}) = -\log(\hat{P}(\vx_{\text{true}})), \]
where \(\hat{P}\) is the estimated posterior PDF, and \(\vx_{\text{true}}\) is the true value. 
A lower log score indicates that the true value has a high probability according to the posterior PDF, implying higher predictive accuracy. Conversely, a higher log score suggests that the true value has a low probability, indicating poor predictive performance, for example, due to too wide uncertainty bounds or biased predictions with small uncertainty bounds. The log score provides a single numerical value for each model parameter that can be used to compare the effectiveness of different probabilistic inversion schemes. To enhance interpretability, we present the results in the pixel domain rather than in terms of the principal components used in the inversion.

\begin{figure}
\centering
\includegraphics[width=1\textwidth]{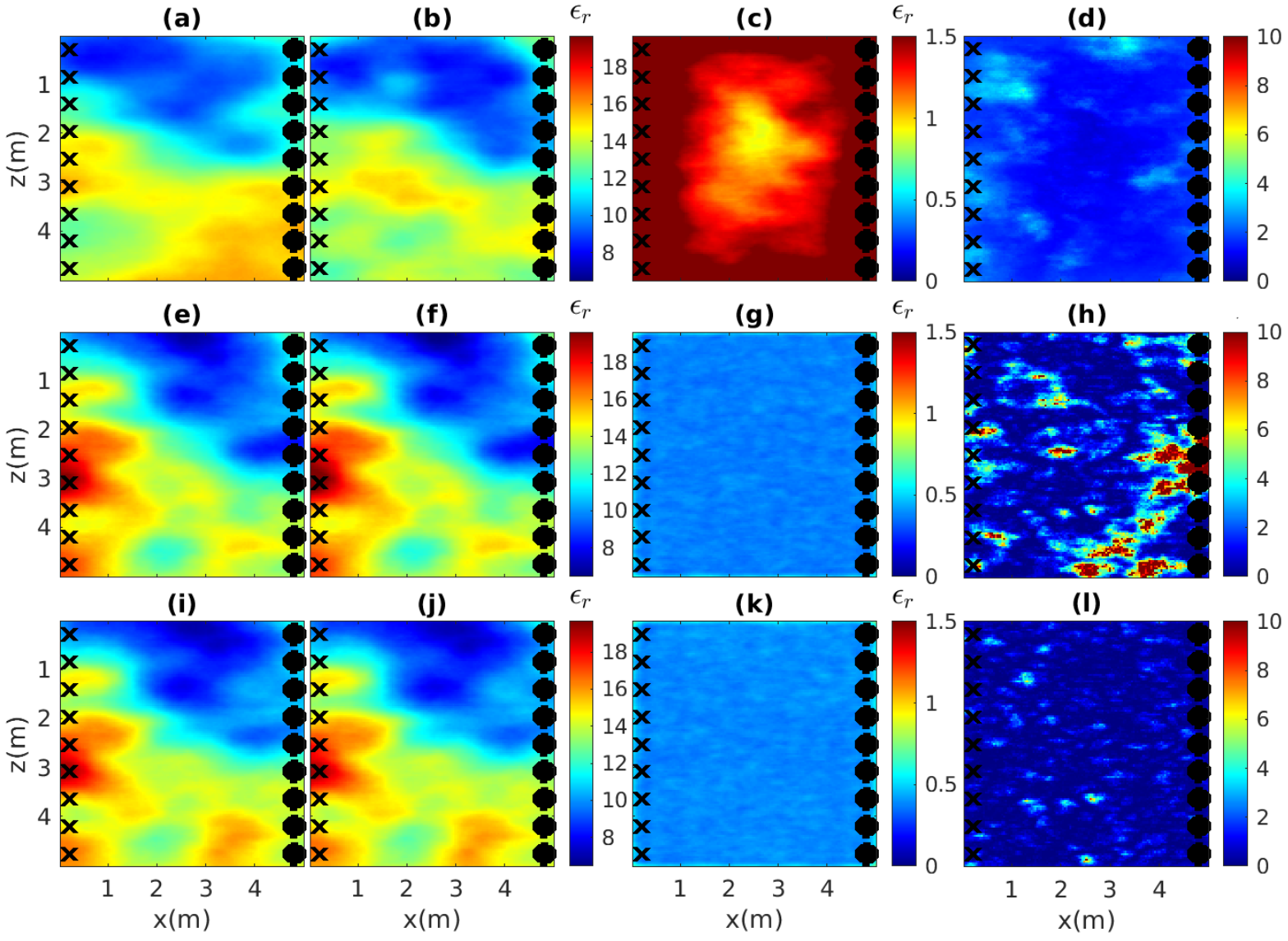}
\caption{\label{FIG_007} (a) Posterior MEAN $\epsilon_r$, (b) MAP $\epsilon_r$, (c) $ \sigma_{\epsilon_r}$ and (d) log score maps evaluated with respect to the true model shown in Figure \ref{FIG_001}(a) as provided by travel time inversion with an eikonal solver. (e-h): as for (a-d), but for surrogate modeling of waveform data inverted by the FBPT scheme. (i-l), as for (a-d), but for waveform data inverted by the PEPT(0) scheme.}
\end{figure}

\begin{figure}
\centering
\includegraphics[width=1\textwidth]{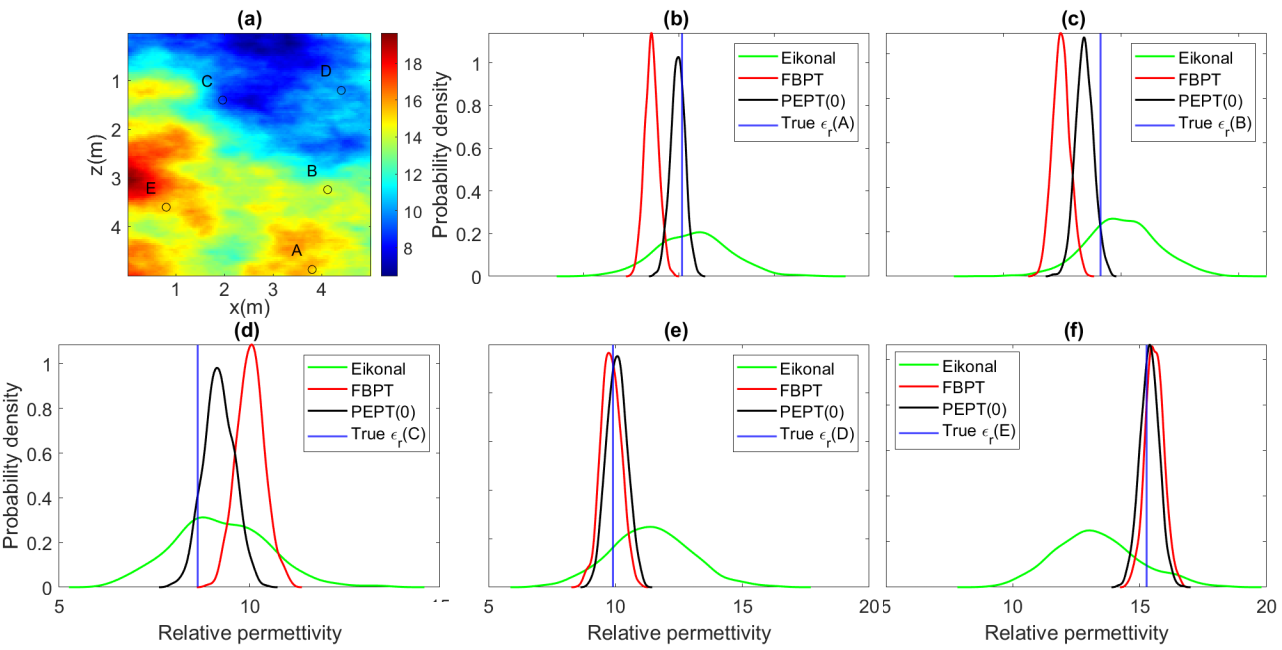}
\caption{\label{IMG_MARG} For selected points A–E distributed across the permittivity distribution (a), (b–f) show the overlay of the corresponding posterior marginals for the eikonal, FBPT, and PEPT(0) inversion schemes.}
\end{figure}

The results for the test permittivity distribution in Fig. \ref{FIG_001}(a) are summarized in Fig. \ref{FIG_007}. It is seen that the eikonal inversion results capture some of the main features of the target distribution (Fig.  \ref{FIG_007}a-d), with a mean value of $\sigma_{\epsilon_r}$ throughout the domain close to 1.55. Conversely, both the FBPT (Fig.  \ref{FIG_007}e-h) and PEPT (Fig.  \ref{FIG_007}i-l) surrogate-based waveform strategies provide much smaller $\sigma_{\epsilon_r}$ (0.37 and 0.43 for the FBPT and PEPT schemes, respectively). However, for the FBPT scheme, this reduction in uncertainty is accommodated by high log score values due to bias introduced by the surrogate modeling. The PEPT strategy, however, with its progressively-improving modeling accuracy in the region of high posterior probability is capable of reducing both uncertainty and log score values compared to the eikonal inversions, thus emerging as the most suitable of the three considered algorithms.
The reduction in bias in PEPT compared to FBPT, as well as its lower uncertainty relative to the eikonal scheme, can be better appreciated by analyzing the performance of the three methods at specific points (Fig. \ref{IMG_MARG}).
To highlight these differences, we evaluate posterior marginals at selected locations across the permittivity distribution (points A-E in Fig. \ref{IMG_MARG}(a)). In Fig. \ref{IMG_MARG}(b-f), we show the marginal posteriors for the three inversion schemes.
The eikonal inversion results in broad distributions, indicating high uncertainty, though the true values consistently retain nonzero probability. In contrast, FBPT significantly reduces uncertainty but introduces noticeable bias, as seen in Fig. \ref{IMG_MARG}(b-d), where the reference values have very low probability, as also shown in the corresponding log score diagram in Fig. \ref{FIG_007}(h). Finally, PEPT achieves both low uncertainty and an unbiased result, with the reference values always exhibiting significant probability at each selected point.  The results for the test permittivity distributions shown in Figs. \ref{FIG_001}(c) and (e) are given in Appendix \ref{Appendix:Results}.

\begin{figure}
\centering
\includegraphics[width=1\textwidth]{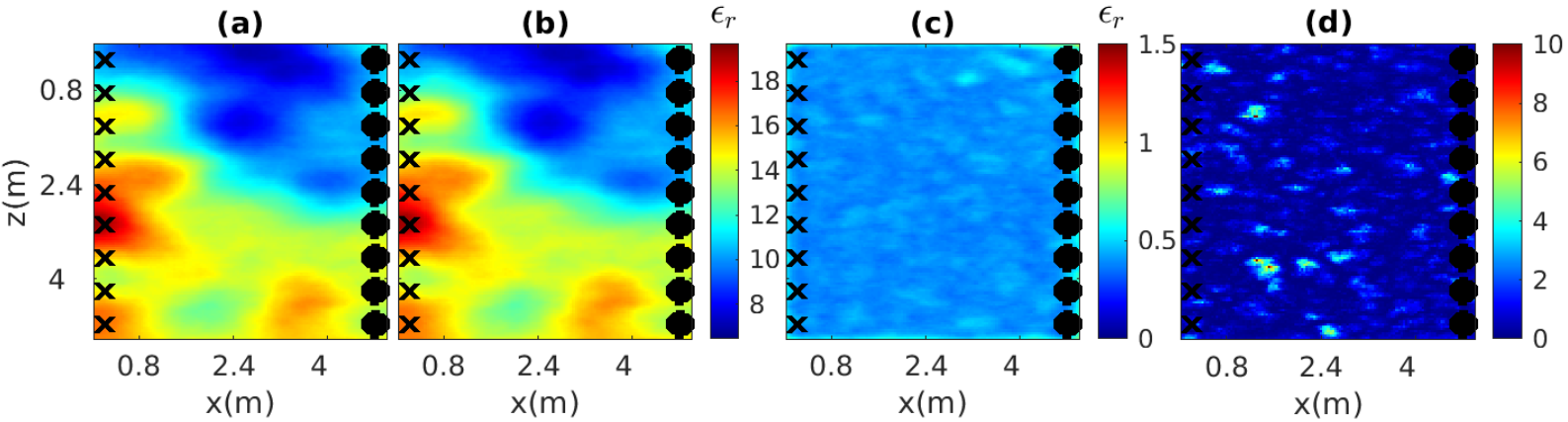}
\caption{\label{IMG_005} (a) Posterior MEAN $\epsilon_r$, (b) MAP $\epsilon_r$, (c) $ \sigma_{\epsilon_r}$ and (d) log score maps evaluated with respect to the true model shown in Figure \ref{FIG_001}(a) using the PEPT(1) scheme.}
\end{figure}

To speed up the inversion process in the PEPT scheme, results from different MCMC runs are combined for components shared between consecutive iterations, with the final samples from one iteration serving as the initial samples for the next. For completeness, we also performed an inversion using the final PEPT model for the distribution in Fig. \ref{FIG_001}(a), starting with samples randomly drawn from the  prior. In the following, we refer to PEPT as combining samples from consecutive iterations as PEPT(0), and to the application of the final surrogate model obtained by applying PEPT(0) to samples from the prior distribution as PEPT(1). The results for PEPT(1) (Fig. \ref{IMG_005}) are very similar to those of PEPT(0) (Fig. \ref{FIG_007}).

Additional metrics of the inversion results in the input and output domains for the tests associated with the permittivity distribution in Fig. \ref{FIG_001}(a) are summarized in Tables \ref{tab:OUTPUT_RESULTS} and \ref{tab:INPUT_RESULTS}.

\begin{figure}
\centering
\includegraphics[width=1\textwidth]{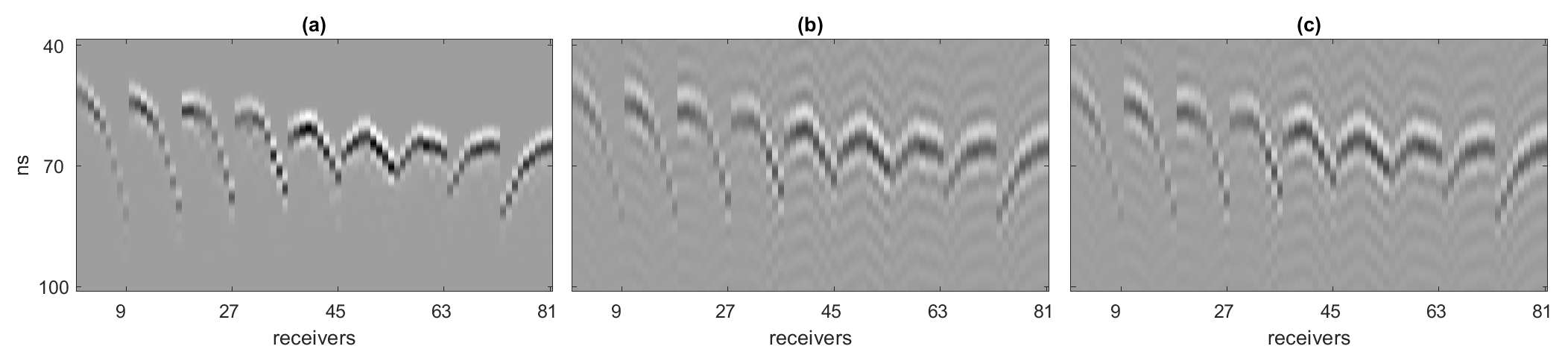}
\caption{\label{FIG_PEPT} (a) Full-bandwidth response for the relative permittivity distribution shown in Fig. \ref{FIG_001}(a).(b) Passband-filtered response of (a) within the 0–144 MHz bandwidth.
(c) Projection of (b) onto the 864 output PCs considered in the PEPT(0) scheme.} 
\end{figure}

A visual comparison of the full-bandwidth, passband-filtered, and projected datasets used in the $4^{th}$ PEPT iteration (Fig. \ref{FIG_PEPT}) demonstrates that the same number of output PCs can more accurately approximate the reference data when associated with a significantly more focused input domain (see Fig. \ref{FIG_FBPT}(c) vs. Fig. \ref{FIG_PEPT}(c)).  Similar to the FBPT strategy, the first two components capture, on average, about 60\% of the total variance across the various frequency bins and minigathers considered here. However, updating the output PCs as we focus in high-posterior regions is crucial for optimally capturing relevant data details. While the relative variance captured by the employed PCs may be similar between FBPT and PEPT, the overall agreement is noticeably improved when PEPT is applied.

Finally, a note on convergence: two of the waveform algorithms considered here, FBPT and PEPT(0), produced chains that achived convergenge for all parameters according to the Gelman-Rubin criterion \citep{gelman1992inference}. Unsurprisingly, the FBPT algorithm achieved convergence in fewer steps due to its larger modeling error covariance matrix. In contrast, the PEPT(1) algorithm exhibited weaker convergence properties, with only one-fifth of the parameters reaching convergence. While longer chains could potentially improve convergence, a more in-depth analysis of this aspect falls outside the primary scope of this study.

\begin{table}[ht]
\centering
\caption{Summary of the inversion results in the data domain for the permittivity distribution in Fig. \ref{FIG_001}(a).}
\begin{tabular}{@{}lcc@{}}
    \toprule
   ALGORITHM: &DATA RMSE MEAN (V/m) & DATA RMSE MAP (V/m)  \\
    \midrule
    EIKONAL & 0.277 & 0.626   \\
    FBPT & 0.393 & 0.380  \\
    PEPT(0) & 0.046 & 0.048 \\
    PEPT(1) & 0.053 & 0.056 \\
    \bottomrule
\end{tabular}
\label{tab:OUTPUT_RESULTS}
\end{table}

\begin{table}[ht]
\centering
\caption{Summary of the inversion results in the input domain for the permittivity distribution in \ref{FIG_001}(a).}
\begin{tabular}{@{}lcccccr@{}}
    \toprule
   ALGORITHM: &RMSE MEAN$_{\epsilon_r}$ & RMSE MAP$_{\epsilon_r}$ &SSIM MEAN$_{\epsilon_r}$ & SSIM MAP$_{\epsilon_r}$ & $\bar{\sigma}_{POST}$ & $\overline{\log(S)}$\ \\
    \midrule
    EIKONAL & 1.44 & 1.50 & 0.64 & 0.61 & 1.54 & 1.72  \\
    FBPT & 0.71 & 0.76 & 0.80 & 0.80 & 0.38 & 2.56  \\
    PBET(0) & 0.37 & 0.38 & 0.82 & 0.81 & 0.44 & 0.48 \\
    PBET(1) & 0.39 & 0.39 & 0.80 & 0.80 & 0.40 & 0.50 \\
    \bottomrule
\end{tabular}
\label{tab:INPUT_RESULTS}
\end{table}

\section{Discussion}
We have introduced a Bayesian FWI scheme that is accelerated by sequential model refinement through progressive frequency bandwidth expansion (PEPT), and applied it to GPR data. Our findings offer compelling evidence for the effectiveness of this approach compared to other approaches  using an eikonal forward solver or training one surrogate model only as in the FBPT approach, yet several key areas warrant further investigation.

 Identifying optimal frequency bands could significantly enhance the robustness of the PEPT scheme. Another important factor is the number of PCs used in each iteration. The computational efficiency and accuracy of the inversion are influenced by the choice of PCs, which in turn depend on the inverted frequency bandwidth and the experimental design. In theory, inclusion of higher frequencies would require more PCs. In this study, we limited the maximum number at 100
 Interestingly, the polynomial complexity of FBPT and PEPT differs markedly when applied to the same input domain (i.e., in the last PEPT iteration), particularly in terms of the required truncation scheme. To reach the relatively modest accuracy reported in this manuscript, FBPT required a maximum polynomial degree of five, with higher degrees proving computationally prohibitive. In contrast, PEPT achieved significantly better accuracy with a maximum degree of only two. While the polynomial degree in FBPT had to be capped due to the associated computational cost, PEPT attained sufficient accuracy well before computational constraints became a limiting factor.
The relative simplicity of the PCE associated with PEPT also results in significantly faster execution times.
 Although the PEPT scheme is less computationally demanding in terms of PCE complexity and could accommodate even more PCs, we limited the input dimensionality to ensure a fair comparison. Future work could explore adaptive approaches that dynamically adjust the number of PCs based on the frequency range and the complexity of the subsurface model, potentially enhancing performance. Another promising direction involves applying an additional PCA to exploit correlations among the currently used PCs under the posterior pdf. If successful, this approach could further accelerate the PEPT scheme, whose computational cost scales linearly with the number of output components.

The size of the training set used for surrogate model construction also plays an important role in determining accuracy. While larger training sets capture more variability, they come at a higher computational cost. Research in how to determine optimal training set sizes and efficient sampling strategies would be beneficial to strike the right balance between accuracy and computational efficiency. Additionally, the way input and output parameters are defined significantly impacts the surrogate model's accuracy and, by extension, the entire inversion process. In cases involving complex geological priors, advanced dimensionality reduction techniques beyond PCA may be required. Deep generative models (DGMs) present a promising tool for this, although they pose challenges for pce-based surrogate modeling. Nevertheless, MCMC strategies that combine the high reconstruction capabilities of DGMs with the accuracy of PCA-PCE surrogate modeling have been proposed for travel-time tomography \cite{meles2024bayesian} and could be explored for waveform data applications.

Although we employed PCE for the surrogate model in this study, other techniques, such as deep learning, could offer alternative avenues for investigation. Our sequential surrogate model refinement approach is driven by frequency expansion, leveraging the nature of the input-output relationship in wave phenomena, where lower frequencies are less sensitive to small-scale variations.

In this study, we did not invert for electrical conductivity and we assumed known source characteristics, which simplified the problem by reducing the number of parameters. In seismic applications, complexity increases significantly due to multiple parameters, such as compressional wave velocity, shear wave velocity, and density, resulting in a greater number of input variables compared to ground-penetrating radar (GPR) scenarios. Nevertheless, our approach of progressively building more accurate surrogates is robust and should be tested across different wave phenomena and acquisition settings, including seismic surface data.


In terms of computational expenses, all the numerical experiments in this paper were performed on a workstation equipped with 16GB of DDR4 RAM and a 3.5GHz Quad-Core processor, running MATLAB on Linux. The three methods we examined - eikonal, FBPT, and PEPT - incurred similar computational costs, averaging around 10 hours. We opted not to conduct an FDTD-based MCMC inversion due to its prohibitively high computational demands \citet{hunziker2019bayesian}. If we had followed the same number of iterations as in the final PEPT strategy, the execution time would have been roughly 200 times longer than that required for a full PEPT inversion. In contrast to FDTD-based inversion, our strategies can be executed on standard computing hardware within a practical timeframe, making them accessible for a variety of applications. Additionally, the algorithm is comparatively simple and can be easily adapted to meet specific user requirements.

\section{Conclusions}

We have introduced a novel strategy for accelerating MCMC inversion of waveform data using surrogate modeling. Our approach employs a sequential, data-driven method to circumvent the challenge of training a single surrogate model capable of capturing the complexity of the input-output relationship across the entire prior pdf.
We leverage the strong relationship between the low-wavenumber and low-frequency components of both parameter space and data domain. The input and output functions are parameterized using appropriate principal component schemes to derive an initial surrogate model with limited coverage but sufficient accuracy.
This preliminary surrogate model serves as a basis for refining modeling accuracy across an expanded bandwidth range through retraining with posterior samples from MCMC targeting limited bandwidths of the data. As the frequency content of the inverted data is expanded, so is the dimension of the input space used to model it, as shorter wavelength structures become more and more relevant in explaining the data.
Through an iterative process, we ultimately obtain a final surrogate model capable of inverting the entire dataset for the definitive MCMC inversion. Unlike travel time inversion via surrogate modeling, where modeling errors are typically small compared to noise, surrogate-based waveform inversion schemes can encounter larger modeling errors, especially for high frequencies. Hence, incorporating these errors in the covariance term of the likelihood is necessary. Our strategy yields significantly better results than those provided by an eikonal solver at comparatively low computational costs, while a surrogate-based waveform modeling approach trained only on samples from the prior pdf leads to biased estimates with misleadingly small uncertainty estimates.


\section*{Data availability}
The data and scripts underlying this manuscript will be shared on reasonable request to the authors.

\appendix
\section{Appendix A}
\setcounter{figure}{0}
\setcounter{equation}{0}
\renewcommand{\thefigure}{A.\arabic{figure}}
\renewcommand{\theequation}{A.\arabic{equation}}
\label{Appendix:PCE}
For this study, we rely on PCE-based surrogate modeling due to its efficiency, flexibility  and ease of deployment \citep{xiu2002wiener,blatman2011adaptive,luethen2021sparse,metivier2020efficient}. 
PCE approximates functions in terms of linear combinations of orthonormal multivariate polynomials $\Psi_{\boldsymbol{\alpha}}$:
\begin{equation}
 \tilde{\mathcal{M}} (\vx ) = \sum_{\boldsymbol{\alpha} \in  \mathcal{A}} a_{\boldsymbol{\alpha}} \Psi_{\boldsymbol{\alpha}}(\vx ),
 \label{PCE_Total}
\end{equation}
where $M$ is the dimension of $\vx $ and  $\mathcal{A}$ is a subset of $\mathbb{N}^{{M}}$ defined by a truncation scheme to be set based on accuracy requirements and available computational resources \citep{xiu2002wiener}. 

Training the coefficients $a_{\ve{\alpha}}$ is computationally unfeasible when the input domain is high-dimensional (the case for a surrogate $\tilde{\mathcal{F}}(\vu)$ of $\mathcal{F}(\vu)$).
To alleviate this computational challenge, we sparse regression techniques to identify and retain only the most relevant polynomial basis terms. However, while such approaches significantly reduce the number of required model evaluations and enable the use of higher polynomial degrees, the fundamental concerns related to the choice of truncation scheme - especially its capacity to capture the model’s complexity - still remain.
In this paper we showed how training the surrogate to approximate the model response only in regions of interest can decrease modeling complexity and thus increase accuracy . 
Once derived, the surrogate response can be evaluated at a negligible cost by direct computation of Eq.~\eqref{PCE_Total} and its accuracy estimated using a validation set or cross-validation techniques \citep{blatman2011adaptive,UQdoc_14_104}.

\section{Appendix B}
\setcounter{figure}{0}
\setcounter{table}{0}
\setcounter{equation}{0}
\renewcommand{\thefigure}{B.\arabic{figure}}
\renewcommand{\thetable}{B.\arabic{table}}
\renewcommand{\theequation}{B.\arabic{equation}}
\label{Appendix:Results}

In the main body of this manuscript, we have presented results using the eikonal, FBPT, and PEPT(0) inversion schemes, along with the corresponding $\epsilon_r$ distribution shown in Fig. \ref{FIG_001}(a). To demonstrate the general character of these results, we provide here the results obtained when considering the permittivity distributions in Figs. \ref{FIG_001}(c) and (e). Note that all the examples presented here share the same prior and training parameters. 
These results are consistent with what is shown in Fig. \ref{FIG_007}. The travel-time inversion results exhibit high uncertainty (Figs. \label{IMG_008}(c) and \label{IMG_009}(c)). This uncertainty decreases dramatically when the FBPT scheme is applied (Figs. \label{IMG_008}(g) and \label{IMG_009}(g)), but these results may imply a significant bias, as indicated in the log score maps (Figs. \label{IMG_008}(h) and \label{IMG_009}(h)). Finally, uncertainty is similarly reduced when the PEPT(0) algorithm is used (Figs. \label{IMG_008}(k) and \label{IMG_009}(k)), but in this case, the bias is drastically reduced, as evidenced by the low log score values (Figs. \label{IMG_008}(l) and \label{IMG_009}(l)). Moreover, the trend observed in the summary of output-input MEAN/MAP fitting in Tables \ref{tab:OUTPUT_RESULTS} and \ref{tab:INPUT_RESULTS} for the first numerical experiment is further confirmed in Tables \ref{tab:OUTPUT_RESULTS_55} to \ref{tab:OUTPUT_RESULTS_91}, which correspond to the numerical tests discussed in this appendix.
\label{Appendix:MoreResults}
\begin{figure}
\centering
\includegraphics[width=1\textwidth]{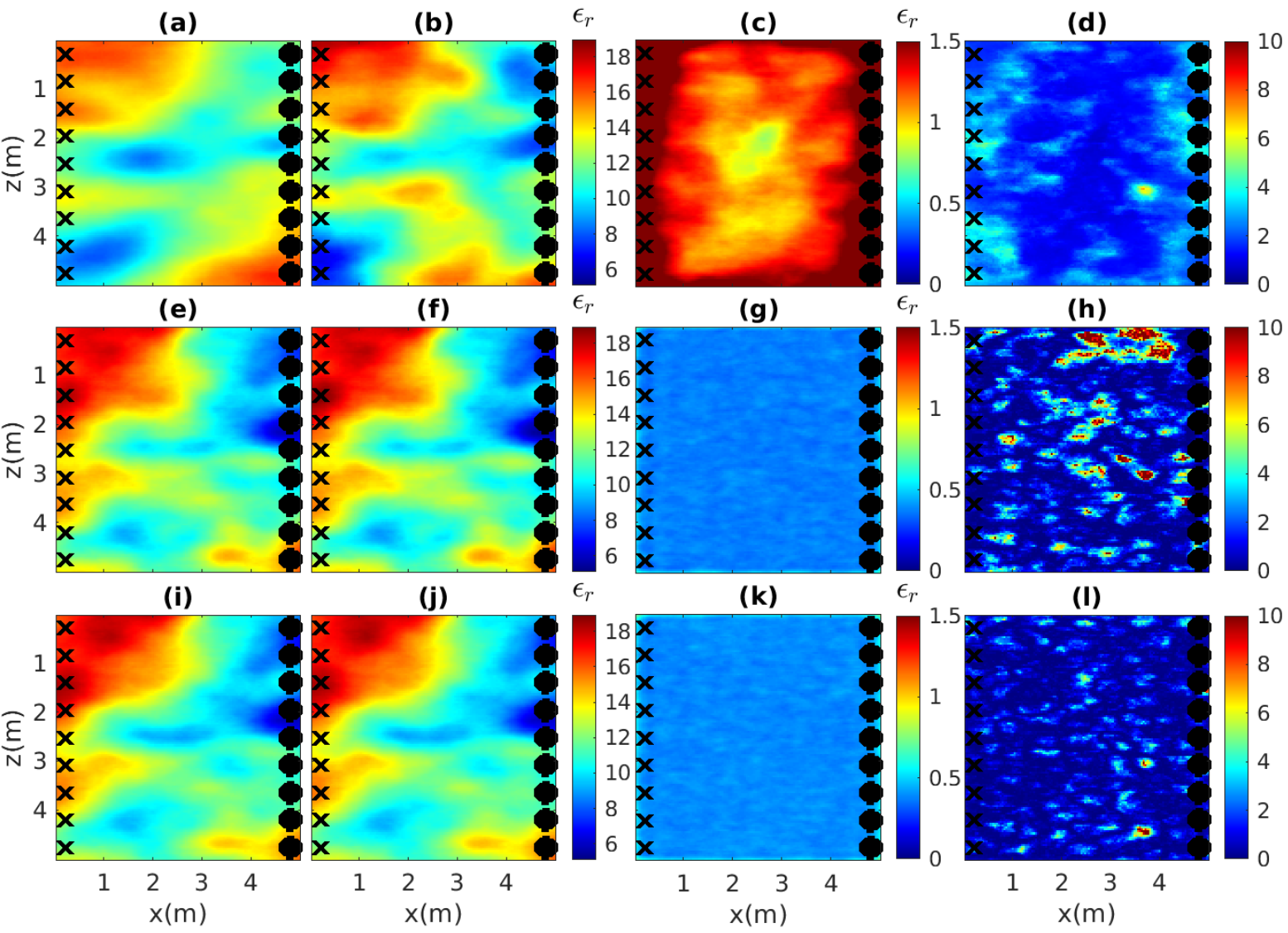}
\caption{\label{IMG_008} (a) Posterior MEAN $\epsilon_r$, (b) MAP $\epsilon_r$, (c) $ \sigma_{\epsilon_r}$ and (d) log score maps evaluated with respect to the true model shown in Figure \ref{FIG_001}(c) as provided by travel time inversion with an eikonal solver. (e-h): as for (a-d), but for surrogate modeling of waveform data inverted by the FBPT scheme. (i-l), as for (a-d), but for waveform data inverted by the PEPT(0) scheme.}
\end{figure}

\begin{table}[ht]
\centering
\caption{Metrics of the inversion results in the data domain for the permittivity distribution in \ref{FIG_001}(c).}
\begin{tabular}{@{}lcc@{}}
    \toprule
   ALGORITHM: &DATA RMSE MEAN (V/m) & DATA RMSE MAP (V/m)  \\
    \midrule
    EIKONAL & 0.292 & 0.301   \\
    FBPT & 0.378 & 0.400  \\
    PEPT(0) & 0.058 & 0.059 \\
    \bottomrule
\end{tabular}
\label{tab:OUTPUT_RESULTS_55}
\end{table}

\begin{table}[ht]
\centering
\caption{Metrics of the inversion results in the input domain for the permittivity distribution in Fig. \ref{FIG_001}(c).}
\begin{tabular}{@{}lcccccr@{}}
    \toprule
   ALGORITHM: &RMSE MEAN$_{\epsilon_r}$ & RMSE MAP$_{\epsilon_r}$ &SSIM MEAN$_{\epsilon_r}$ & SSIM MAP$_{\epsilon_r}$ & $\bar{\sigma}_{POST}$ & $\overline{\log(S)}$\ \\
    \midrule
    EIKONAL & 1.87 & 1.72 & 0.63 & 0.62 & 1.35 & 1.98  \\
    FBPT & 0.62 & 0.66 & 0.78 & 0.78 & 0.37 & 1.74  \\
    PBET(0) & 0.45 & 0.46 & 0.79 & 0.79 & 0.39 & 0.67 \\
    \bottomrule
\end{tabular}
\label{tab:INPUT_RESULTS_55}
\end{table}

\begin{figure}
\centering
\includegraphics[width=1\textwidth]{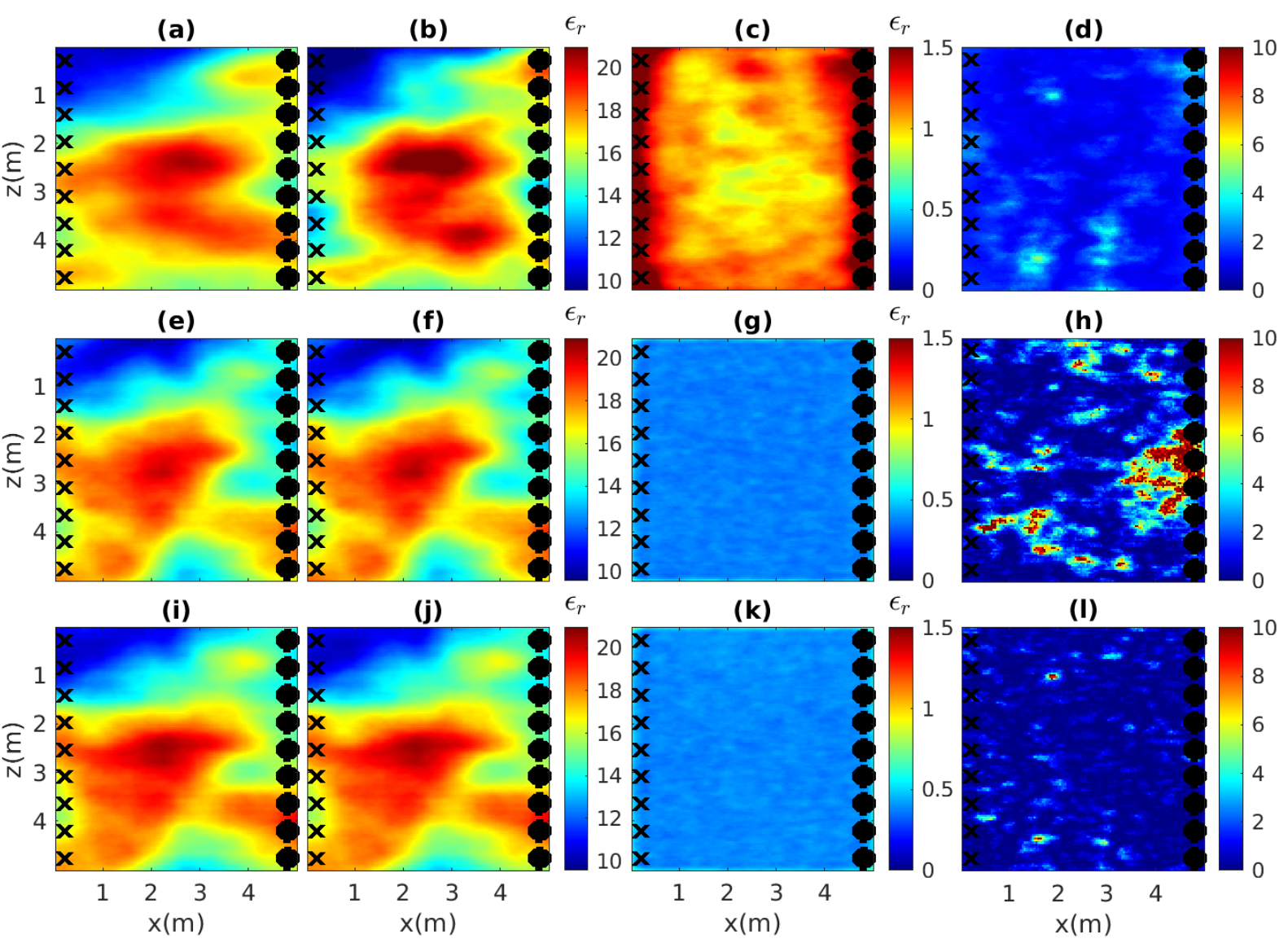}
\caption{\label{IMG_009} (a) Posterior MEAN $\epsilon_r$, (b) MAP $\epsilon_r$, (c) $ \sigma_{\epsilon_r}$ and (d) log score maps evaluated with respect to the true model shown in Figure \ref{FIG_001}(e) as provided by travel time inversion with an eikonal solver. (e-h): as for (a-d), but for surrogate modeling of waveform data inverted by the FBPT scheme. (i-l), as for (a-d), but for waveform data inverted by the PEPT(0) scheme.}
\end{figure}

\begin{table}[ht]
\centering
\caption{Metrics of inversion results in the data domain for the permittivity distribution in Fig. \ref{FIG_001}(e).}
\begin{tabular}{@{}lcc@{}}
    \toprule
   ALGORITHM: &DATA RMSE MEAN (V/m) & DATA RMSE MAP (V/m)  \\
    \midrule
    EIKONAL & 0.201 & 0.267   \\
    FBPT & 0.652 & 0.612  \\
    PEPT(0) & 0.044 & 0.048 \\
    \bottomrule
\end{tabular}
\label{tab:OUTPUT_RESULTS_91}
\end{table}

\begin{table}[ht]
\centering
\caption{Metrics of the inversion results in the input domain for the permittivity distribution in Fig. \ref{FIG_001}(e).}
\begin{tabular}{@{}lcccccr@{}}
    \toprule
   ALGORITHM: &RMSE MEAN$_{\epsilon_r}$ & RMSE MAP$_{\epsilon_r}$ &SSIM MEAN$_{\epsilon_r}$ & SSIM MAP$_{\epsilon_r}$ & $\bar{\sigma}_{POST}$ & $\overline{\log(S)}$\ \\
    \midrule
    EIKONAL & 1.01 & 1.51 & 0.72 & 0.68 & 1.18 & 1.46  \\
    FBPT & 0.69 & 0.69 & 0.80 & 0.80 & 0.37 & 2.31  \\
    PBET(0) & 0.34 & 0.35 & 0.82 & 0.82 & 0.39 & 0.41 \\
    \bottomrule
\end{tabular}
\label{tab:INPUT_RESULTS_91}
\end{table}


\bibliographystyle{apalike}  
\bibliography{paper}

\end{document}